\documentclass[prx,twocolumn,superscriptaddress,floatfix,longbibliography, 10pt]{revtex4-2}

\usepackage[english]{babel}
\usepackage[utf8]{inputenc}
\usepackage[T1]{fontenc}
\usepackage{braket,color}
\usepackage{graphicx,enumerate,verbatim,bbold}
\usepackage{amsmath,amssymb,amsthm,float,mathrsfs}
\usepackage{dsfont}
\usepackage[normalem]{ulem}
\usepackage{lmodern}
\usepackage{multirow}
\usepackage{bbm}
\usepackage{physics}
\usepackage{bm}
\usepackage[normalem]{ulem}
\usepackage[caption=false]{subfig}
\usepackage{dcolumn}
\newcommand{\bqa}{\begin{eqnarray}}
\newcommand{\eqa}{\end{eqnarray}}

\newcommand{\be}{\begin{equation}}
\newcommand{\ee}{\end{equation}}

\begin{document}
\title{Quantum control without quantum states}

\author{Modesto Orozco-Ruiz}
\affiliation{Physics Department, Blackett Laboratory, Imperial College London, Prince Consort Road, SW7 2AZ, United Kingdom}
\author{Nguyen H. Le}
\affiliation{Physics Department, Blackett Laboratory, Imperial College London, Prince Consort Road, SW7 2AZ, United Kingdom}
\author{Florian Mintert}
\affiliation{Physics Department, Blackett Laboratory, Imperial College London, Prince Consort Road, SW7 2AZ, United Kingdom}

\begin{abstract}
We show that combining ideas from the fields of quantum invariants and of optimal control can be used to design optimal quantum control solutions without explicit reference to quantum states. The states are specified only implicitly in terms of operators to which they are eigenstates. The scaling in numerical effort of the resultant approach is not given by the typically exponentially growing effort required for the specification of a time-evolved quantum state, but it is given by the effort required for the specification of a time-evolved operator.
For certain Hamiltonians, this effort can be polynomial in the system size.

We describe how control problems for state preparation and the realization of propagators can be formulated in this approach, and we provide explicit control solutions for a spin chain with an extended Ising Hamiltonian.
The states considered for state-preparation protocols include eigenstates of Hamiltonians with more than pairwise interactions, and these Hamiltonians are also used for the definition of target propagators. The cost of describing suitable time-evolving operators grows only quadratically with the system size, allowing us to construct explicit control solutions for up to $50$ spins.
While sub-exponential scaling is obtained only in special cases, we provide several examples that demonstrate favourable scaling beyond the extended Ising model.

\end{abstract} 
\maketitle

\section{Introduction}

The availability of well-controlled quantum systems offers the prospect towards a broad range of applications~\cite{Koch2022}, including quantum simulation \cite{johnson2014quantum, marcos2013superconducting} and quantum computation \cite{ladd2010quantum, horowitz2019quantum, gyongyosi2019survey}.
The exponential growth of the Hilbert space in such systems suggests that even moderately small quantum systems could be employed to achieve tasks that would be impractical to perform classically~\cite{ladd2010quantum,buluta2009quantum}.
While the benefits of this scaling have inspired tremendous efforts towards the development of quantum technological devices \cite{gyongyosi2019survey, kim2023evidence, wu2021strong, zhong2020quantum, madsen2022quantum}, the scaling also has disastrous implications on our abilities to simulate the dynamics of quantum systems~\cite{daley2022practical}.

Most attempts to devise control schemes of quantum systems are thus restricted to exact analyses performed on small systems \cite{Khaneja2005,schulte-optimal_2005} with potential extensions to larger system via approximate techniques \cite{chetcuti_perturbative_2020,doria2011optimal}, 
specific systems that admit efficient descriptions \cite{PhysRevA.81.040303,le_scalable_2023},
or, to analyses that are approximate to start with~\cite{castro2012controlling}.
While this exponential scaling seems an insurmountable obstacle in general, there are specific instances of quantum states that can be described with an effort that scales more favourably than exponentially \cite{verstraete2008matrix,carleo_solving_2017}. 

A very efficient, even though implicit way of describing some quantum states is given in terms of the operator (or operators) to which a state is an eigenstate \cite{gottesman1997stabilizer}.
There is, for example, a large variety of quantum many-body Hamiltonians with highly complicated ground states \cite{raussendorf2003measurement, levin2005string, Kitaev_2003}. The Hamiltonian itself is specified in terms of the interaction geometry, the type of interaction mechanism and some parameters specifying quantities like interaction strengths.
For many systems, the number of parameters required for this specification scales polynomially with the number of interacting objects; the description is thus {\em efficient}.
The ground state of many such Hamiltonians, however, often lacks an efficient representation, {\it i.e.} a representation specified with a polynomial number of parameters.
Implicitly, however, the ground state is specified via the Hamiltonian.
This implicit description of states is foundational to the stabilizer formalism~\cite{Aaronson2004} and to adiabatic quantum computation~\cite{albash2018adiabatic}. 

Building on this concept of implicit state characterization, this paper introduces a framework for optimal control that avoids explicit construction of quantum states, while not being limited to adiabatic time scales. Formulating an optimal quantum control formalism in terms of operators, we demonstrate a method that can help solving control problems that are computationally intractable using traditional state-vector-based approaches, and that does not require an approximate treatment.

The underlying idea is inspired by the framework of quantum invariants \cite{lewis1969exact, kaushal1993dynamical, korsch1979}, {\it i.e.} hermitian operators ${\cal I}(t)$ satisfying the equation of motion
\be
\frac{\partial{\cal I}}{\partial t}+i[H(t),{\cal I}(t)]=0\ ,
\label{eq:eomi}
\ee
in the Schr\"odinger picture.

Each such invariant has instantaneous eigenvectors, {\it i.e.} generally time-dependent vectors satisfying the eigenvalue relation

\begin{align}
\label{eq:invariant_def}
    \mathcal{I}(t)\ket{\Psi(t)}=\gamma\ket{\Psi(t)} \ .
\end{align}

Due to the unitary dynamics resultant from Eq.~\eqref{eq:eomi}, any eigenvalue $\gamma$ of $\mathcal{I}(t)$ is time-independent.
Since every non-degenerate, instantaneous eigenvector specifies a solution of the time-dependent Schr\"odinger equation with the time-dependent Hamiltonian $H(t)$~\cite{lewis1969exact}, invariants are commonly used tools in optimal quantum control~\cite{guery2019shortcuts}. 

Existing use of quantum invariants in optimal quantum control is based on the identification of analytic time-dependences that make an operator satisfy Eq.~\eqref{eq:eomi} for a certain type of Hamiltonian, such as the driven harmonic oscillator Hamiltonian and the Hamiltonian of a single driven spin \cite{chen2010fast, Torrontegui2014, Simsek2021quantumcontrolmulti}.
Once such an analytic form is identified, the resultant invariant provides very elegant means of engineering a time-dependent Hamiltonian that induces desired dynamics.

Finding such analytic expressions is practical only for particularly simple cases, and it does seem elusive to find an analytic invariant for quantum systems with a large number of interacting subsystems. The goal of this paper is thus to develop an approach that benefits from some aspects of quantum invariants, but that does not rely on the existence of an analytic form of an invariant.
This approach will combine elements from optimal control based on quantum invariants \cite{guery2019shortcuts} and from numerical quantum control with state vectors or propagators~\cite{Koch2022,peirce1988optimal}.
It relies on iterative numerical refinement of a time-dependent Hamiltonian, but instead of the commonly employed numerical integration of the Schr\"odinger equation, the present approach is based on numerical integration of the equation of motion of quantum invariants, {\it i.e.} Eq.~\eqref{eq:eomi}.
Since such numerical integration does not require the existence of analytic solutions, this approach can indeed be applicable to interacting quantum systems. 

Instead of defining a control target for state vectors or propagators, control targets are defined for operators ${\cal I}(t)$, satisfying the equation of motion of quantum invariants Eq.~\eqref{eq:eomi}.
This avoids reference to quantum states with corresponding exponential effort, and, as shown with explicit examples, there are interacting quantum systems for which the description of the time-dependent operator $\mathcal{I}(t)$ does {\em not} require an exponential effort.

\section{
Implicit quantum control}
\label{sec:invariantcontrol}

Instead of numerically propagating the Schr\"odinger equation explicitly for a time-dependent state vector, the proposed approach relies on numerically propagating Eq.~\eqref{eq:eomi} for an operator ${\cal I}(t)$. Following the eigenvalue relation Eq.~\eqref{eq:invariant_def}, this provides all information on the desired state vector, but as long as ${\cal I}(t)$ is not diagonalised, this information remains implicit.

In this section we show how to define a control target ${\cal I}_T$ for an operator ${\cal I}(t)$ following the dynamics of Eq.~\eqref{eq:eomi}, such that numerical pulse shaping techniques can be applied to Eq.~\eqref{eq:eomi},
and such that the resultant Hamiltonian induces desired dynamics in the underlying Schr\"odinger equation.

While translating a control problem from the Schr\"odinger equation to the corresponding  Liouville-von Neumann equation, {\it i.e.} Eq.~\eqref{eq:eomi}, does in general not provide any improvement in efficiency, there are indeed instances such as Eqs.~\eqref{eq:ham} and \eqref{eq:transLie} discussed in more detail below, and a series of further examples listed in Sec.~\ref{sec:scaling} for which this translation results in an exponential improvement.
 
\subsection{Operator expansion}
\label{sex:operatorexpansion}

The crucial aspect to determine whether the present approach is more efficient than approaches based on state vectors is the dimension of the space in which the dynamics of ${\cal I}(t)$ following Eq.~\eqref{eq:eomi} takes place.

Since there is a multitude of different operators with the same ground-state, there is no unique choice for the initial condition ${\cal I}(0)$ to represent a given initial state $\ket{\Psi(0)}$,
and this freedom can be used to find an initial condition ${\cal I}(0)$ that allows for a particularly efficient propagation.

Formalization of this statement requires the definition of system Hamiltonian. In the following, it is considered to be of the form

\be
H(t)=\sum_{j=1}^{d_0}h_j(t){\mathfrak h}_j\ ,
\label{eq:Hamiltonianexpand}
\ee
spanned by a fixed set of $d_0$ operators ${\mathfrak h}_j$ with scalar expansion coefficients $h_j(t)$.
Some of those can be time-independent and fixed (as in a drift Hamiltonian), but some should be time-dependent and tuneable (as in a control Hamiltonian).

The Ansatz
\be
{\cal I} (t) =\sum_{j=1}^d a_j(t){\mathfrak a}_j
\label{eq:Iexpand}
\ee
for the operator ${\cal I}(t)$ to be propagated with a set of $d$ time-independent operators ${\mathfrak a}_j$ and time-dependent scalar expansion coefficients 
$a_j(t)$ is sufficient to solve Eq.~\eqref{eq:eomi} for any choice of the time-dependent coefficients $h_j(t)$, if the commutation relations
\be
[{\mathfrak a}_j,{\mathfrak h}_k] = -i \sum_{l=1}^d\lambda_{lj}^k{\mathfrak a}_l\ ,
\label{eq:structure}
\ee
with real coefficients $\lambda_{lj}^k$ are satsfied for all $k\in[1,d_0]$ and $j\in[1,d]$.

An Ansatz given by Eq.~\eqref{eq:Iexpand} that does not lead to the closed commutator relations of Eq.~\eqref{eq:structure} can be salvaged with an increase of the set of operators ${\mathfrak a}_j$, {\it i.e.} and increase of $d$.
Given a set of $d$ operators ${\mathfrak a}_j$ with a pair of operators ${\mathfrak a}_{j_0}$ and ${\mathfrak h}_{k_0}$ such that the commutator $[{\mathfrak a}_{j_0},{\mathfrak h}_{k_0}]$ can not be expanded in terms of the set of $d$ operators ${\mathfrak a}_j$, one can define ${\mathfrak a}_{d+1}=i[{\mathfrak a}_{j_0},{\mathfrak h}_{k_0}]$ as additional element, increasing the value of $d$ by $1$.
In any finite-dimensional system, this process can be continued until Eq.~\eqref{eq:structure} is satisfied for all pairs $\mathfrak{a}_j$, $\mathfrak{h}_k$.

In practice, this construction can be initiated with a single element $\mathfrak a_1$, {\it i.e.} with $d=1$, and the eventual value of $d$ depends on the choice of initialization.
Since the effort to integrate Eq.~\eqref{eq:eomi} grows with increasing $d$, it is desirable to find small sets of operators $\mathfrak{a}_j$ with closed commutator relationships, but even with suitable choices, $d$ will typically be larger than $d_0$.

\subsection{Lie algebra}\label{sec:lie_alg}

The construction of the set of operators $\{\mathfrak{a}_j\}$ discussed above is a slight variant of the construction of a Lie-algebra,
{\it i.e.} a set of $d$ operators $\mathfrak{h}_j$ with a closed commutator relationship $[{\mathfrak h}_j,{\mathfrak h}_k] = -i \sum_{l=1}^d\lambda_{lj}^k{\mathfrak h}_l$ for all $j,k\in[1,d]$.
Similarly to the construction above, it can be initiated with a set of $d_0$ operators with $d_0<d$ and the construction is completed when all commutation relations are closed.

Crucially, the propagator induced by the time-dependent Hamiltonian $H(t)$ (Eq.\eqref{eq:Hamiltonianexpand}) is \textit{not} necessarily of the form $U(t)=\exp(-i\sum_{j=1}^{d_0}b_j(t)\mathfrak{h}_j)$ with time-dependent scalar functions $b_j(t)$ and the terms $\mathfrak{h}_j$ ($j\in[1,d_0]$) in the system Hamiltonian, 
but typically also terms with $j>d_0$ of the Lie algebra contribute.
In turn, a propagator of the form $U=\exp(-iK)$ with an operator $K$ that cannot be expanded in terms of the basis elements of the Lie algebra ({\it i.e.} $\mathfrak{h}_j$, $j\in[1,d]$) cannot be realized with the dynamics induced by $H(t)$ in Eq.~\eqref{eq:Hamiltonianexpand} and is thus considered not reachable \cite{d2021introduction}. 

In the explicit examples Sec.~\ref{sec:statetransfer} and \ref{sec:realization_propagators} the time-evolving operator will be expanded in terms of the Lie algebra generated by the operators $\mathfrak{h_j}$ with $j\in[1,d_0]$.
In order to find small sets of operators $\{{\mathfrak a}_j\}$ it can, however, also be beneficial to consider situations beyond Lie algebras, as exemplified in Sec.~\ref{sec:scaling}.

\subsection{Equations of motion}

With the explicit expansion in terms of the sets of operators $\{{\mathfrak a}_j\}$ and $\{{\mathfrak h}_k\}$, 
the equation of motion Eq.~\eqref{eq:eomi} reads~\cite{iachello2006lie}

\begin{align}
\frac{\partial{\cal I}}{\partial t}  = i\sum_{j,k} a_j(t) h_k(t) \comm{\mathfrak{a}_j}{\mathfrak{h}_k} = \sum_{j,k,l} a_j(t) h_k(t) \lambda_{lj}^k \mathfrak{a}_l \ .
\end{align}

Since the set of operators $\mathfrak{a}_j$ is linearly independent, this implies the differential equation

\begin{align}
\dot a_l(t)=\sum_{j,k}\lambda_{lj}^ka_j(t)h_k(t)\ ,
\end{align}
that can be written compactly as

\begin{align}
    \dot{\bf a}(t)=K(t) {\bf a}(t)\ ,
\label{dynamics_aj}
\end{align}
with the vector ${\bf a}$ and a matrix $K$ with elements 

\begin{align} 
\label{eq:Kmatrix2}
    K_{lj} (t) =\sum_k h_k(t) \lambda_{lj}^k\ .
\end{align} 

The hermitian operators $\mathfrak{a}_j$ can always be chosen to be mutually orthonormal, such that Eq.~\eqref{eq:structure} implies
\begin{align}
\lambda_{lj}^k=
i \tr([\mathfrak{a}_j,\mathfrak{h}_k]\mathfrak{a}_l)\ .
\end{align}
Invariance of the trace under cyclic permutation implies
\begin{align} \lambda_{lj}^k=i\tr([\mathfrak{a}_l,\mathfrak{a}_j]\mathfrak{h}_k)=-\lambda_{jl}^k\ .
\end{align}
In this case, the matrix $K$ is real and anti-symmetric \cite{Torrontegui2014} such that the induced dynamics is orthogonal, {\it i.e.} unitary and real.

\subsection{Formulation of the control problem}

If formulated in terms of Eq.~\eqref{dynamics_aj} instead of the underlying Schr\"odinger equation,
the dimension of the dynamics underlying the control problem is given by the size $d$ of the set of operators $\{{\mathfrak a}_j\}$, and not by the dimension of the underlying Hilbert space.

In general, $d$ can be quadratically larger than the size of the Hilbert space, but as exemplified in the following, there are several instances of practical importance in which $d$ is much smaller. The goal of the following discussion is to benefit from this reduced dimensionality for optimal control problems.

While typically control problems are defined in terms of a target state or a target gate, the present framework requires a control target for the operator ${\cal I}$, and defining this requires some care.
In the following, we will discuss this for the problems of state preparation and for the realization of propagators.

\subsubsection{State preparation}
\label{subsec:diabatictransitions}

The problem of state preparation is a natural application of the present framework, because Hamiltonians can typically be specified efficiently, whereas the explicit specification of their eigenstates is often not efficient.

With a control problem defined in terms of a target state, the target functional is naturally defined in terms of this state. 
The present approach leaves much more freedom, since there are many operators ${\cal I}_T$ that have a given target state as ground state.
There are thus many conceivable choices for ${\cal I}_T$ and a good choice is a central part of the control problem within the present approach.

For the problem of finding a time-dependent Hamiltonian such that the system evolves from the ground-state of an initial operator $H_I$ to the ground-state of a final operator $H_T$, it seems tempting to define the initial condition ${\cal I}(0)=H_I$ and the control target ${\cal I}_T=H_T$. 
However, since the spectrum of ${\cal I}(t)$ is time-independent, this would necessarily imply the restriction that $H_I$ and $H_T$ have the same spectrum.

Moreover, if the operators $\mathfrak{h}_j$ in the system Hamiltonian (Eq.\eqref{eq:Hamiltonianexpand}) do not generate the maximal Lie algebra, {\it i.e.} an algebra that contains a complete set of operators apart from the identity, then there are unitary propagators that cannot be realised with the system Hamiltonian, as discussed in the penultimate paragraph of Sec.~\ref{sec:lie_alg}.
That means that even if $H_I$ and $H_T$ have the same spectrum, it is not necessarily possible to realise the dynamics such that an operator ${\cal I}(t)$ with the initial condition ${\cal I}(0)=H_I$ evolves towards the final value ${\cal I}(T)=H_T$;
{\it i.e.} the target ${\cal I}_T=H_T$ is not reachable  (explicit examples demonstrating that specific targets can be reached are discussed in Secs.~\ref{sec:targetGHZ} and 
\ref{sec:targetcluster}).

It is thus necessary to impose the requirements that
\begin{itemize}
\item[(i)] the initial condition ${\cal I}(0)$ needs to be specified such that the initial state $\ket{\Psi_I}$ (an eigenvector of $H_I$) is an eigenstate of ${\cal I}(0)$ with a non-degenerate eigenvalue $\gamma$; and that
\item[(ii)] the control target ${\cal I}_T$ needs to be specified such that the target state $\ket{\Psi_T}$ (an eigenvector of $H_T$) is an eigenstate of ${\cal I}_T$ with the same eigenvalue $\gamma$,
\item[(iii)] the control target needs to be reachable with the dynamics induced by the system Hamiltonian $H(t)$, \textit{i.e.}, the dynamics following Eq.~\eqref{dynamics_aj}.
\end{itemize}

In attempting to satisfy these requirements it can be helpful to consider the following aspects: 

For (i): In most control problems, the initial operator $H_I$ would be a non-interacting operator, {\it i.e.} a sum of single-qubit terms (or, more generally single-particle terms) without any interactions.
As such, also the initial condition ${\cal I}(0)$ can be chosen to be non-interacting. 
As long as the single-qubit parts of ${\cal I}(0)$ are non-degenerate, such an operator has a unique, non-degenerate ground-state, and the associated eigenvalue is given by the sum of all the single-qubit ground-state eigenvalues.
In cases in which the initial operator ${\cal I}(0)$ is interacting, it might be challenging to verify the non-degeneracy of the ground state. There is certainly no unique recipe to do this, but one option would be to show unitary equivalence of this operator with a non-degenerate, non-interacting operator.

For (ii): There is a general approach to numerically construct a reachable control target that is applicable if the set of operators $\{\mathfrak{a}_j\}$ is taken to coincide with the Lie algebra generated by the operators $\mathfrak{h}_j$ defining the system Hamiltonian in Eq.~\eqref{eq:Hamiltonianexpand}.
Given any initial condition ${\cal I}(0)$, one can numerically simulate the dynamics of ${\cal I}(t)$ under a Hamiltonian that evolves adiabatically from $H_I$ to $H_T$, as demonstrated  in Sec.~\ref{sec:hd}.
In practice, this implies numerical integration of Eq.~\eqref{dynamics_aj}, with an effort that scales with $d$ and not with the exponentially large dimension of the system Hilbert space.
The final operator ${\cal I}_T$ obtained in this fashion necessarily has the same spectrum as ${\cal I}(0)$ because of unitarity of the dynamics, its ground state coincides with the ground state of $H_T$ because of adiabaticity, and it is reachable with the dynamics induced by the system Hamiltonian.

For (iii): Given Eq.~\eqref{dynamics_aj}, the question of reachability of an operator ${\cal I}_T$ starting from an operator ${\cal I}(0)$ is equivalent to the question of reachability of a vector ${\bf a}^T$ (such that ${\cal I}_T=\sum_ja_j^T\mathfrak{a}_j$) starting from a vector ${\bf a}(0)$ (such that ${\cal I}(0)=\sum_ja_j(0)\mathfrak{a}_j$) given the time-dependent operators $K(t)$ (Eq.~\eqref{eq:Kmatrix2}). The question of reachability at hand is thus equivalent to standard reachability of vectors with a linear differential equation, as in the case of state control with the Schr\"odinger equation \cite{schirmer_criteria_2002}.

The requirements (i), (ii) and (iii) do not result in a unique choice of initial condition ${\cal I}(0)$ and target ${\cal I}_T$, and this freedom of choice can be used to select an initial condition ${\cal I}(0)$ that can be expanded in a set of operators $\{\mathfrak{a}_j\}$ such that the commutation relations Eq.~\eqref{eq:structure} are closed even though $d$ is small.

Some of the examples discussed in Sec.~\ref{sec:statetransfer}, will rely on the numerical construction of control targets with simulated adiabatic dynamics.
While such a numerical construction is a generally applicable approach, there are also instances in which suitable control targets can be defined analytically as exemplified in Secs.~\ref{sec:targetGHZ} and \ref{sec:targetcluster}.

\subsubsection{State fidelity}\label{Sec:fidelity}

Given a consistent definition for initial condition ${\cal I}(0)$ and control target ${\cal I}_T$,
a generic choice for objective function is the infidelity
\be
{\cal J}=1-\frac{\mbox{tr}({\cal I}(T)\, {\cal I}_T)}{\mbox{tr}({\cal I}_T^2)}\ ,
\label{eq:J}
\ee
where $T$ is the final time at which the goal of the control problem is intended to be reached.
Since the infidelity is readily expressed as
\be
{\cal J}=1-\frac{\bm{a}(T)\cdot \bm{a}_T}{\|\bm{a}_T\|^2}\ ,
\label{eq:J2}
\ee
with the vector $\bm{a}_T$ of the control target ${\cal I}_T$, the effort to evaluate the infidelity scales with the dimension $d$ and not with the dimension of system's Hilbert space.

If ${\cal I}(T)$ coincides exactly with the control target ${\cal I}_T$, then it is ensured that the system state coincides with the desired non-degenerate eigenstate of ${\cal I}_T$, {\it i.e.} with the target state.
Since both state infidelity and the infidelity defined in Eq.~\eqref{eq:J} are continuous functions, continuity implies that in the limit of vanishing infidelity ${\cal J}$ for the non-degenerate operator ${\cal I}(T)$ also the infidelity of the realized quantum state approaches its ideal value.
In practice, a more quantitative estimate is desirable to assess the quality of an optimal control solution.

In order to obtain such an estimate that will depend on the spectrum of $H_T$, it is helpful to express the final state $\ket{\Psi(T)}$ as
\be
\ket{\Psi(T)}=\braket{\Psi_T}{\Psi(T)}\ket{\Psi_T} + P_e\ket{\Psi(T)}\ ,
\ee
with the target state $\ket{\Psi_T}$, {\it i.e.} the ground state of $H_T$, and the projector $P_e$ onto the space spanned by the excited states of $H_T$.

The expectation value of $H_T$ with respect to $\ket{\Psi(T)}$ reads 
\be
\bra{\Psi(T)}H_T\ket{\Psi(T)} = \vert \braket{\Psi_T}{\Psi(T)}\vert^2 E_0 + \bra{\phi}H_T\ket{\phi}\ ,
\ee
with $\ket{\phi}=P_e\ket{\Psi(T)}$ and the ground state energy $E_0$ of $H_T$.
The norm of the vector $\ket{\phi}$ is given in terms of the state fidelity $F=|\braket{\Psi_T}{\Psi(T)}|^2$ via the relation
\be
\braket{\phi}{\phi}=1-F \ .
\ee
With the inequality
\be
\frac{\bra{\phi}H_T\ket{\phi}}{\braket{\phi}{\phi}}\ge E_1\ ,
\ee
in terms of the energy $E_1$ of the first excited state of $H_T$, the state fidelity is bounded by
\be
1-F\le\frac{\bra{\Psi(T)}H_T\ket{\Psi(T)}-E_0}{E_1-E_0}\ .
\label{bound_infidelity}
\ee

The state infidelity $1-F$ can thus be bounded in terms of the ground state energy $E_0$, the first excited state energy $E_1$ of the Hamiltonian $H_T$
and the expectation value $\bra{\Psi(T)}H_T\ket{\Psi(T)}$.
At first sight, it might seem that evaluating this expectation value requires explicit construction of the state vector $\ket{\Psi(T)}$.
Since, however, the state vector $\ket{\Psi(t)}=U(t)\ket{\Psi(0)}$ follows the unitary dynamics $U(t)$ induced by the system Hamiltonian, the expectation value can be evaluated in terms of the initial state $\ket{\Psi(0)}$, and a time-evolved operator $U^\dagger(T)H_TU(T)$. The latter can be obtained by backward-propagation of Eq.~\eqref{eq:eomi},
and for any operator $H_T$ expanded in terms of the operators $\{\mathfrak{a}_j$\} ($j\in[1,d]$) this can be done with same effort as required for propagation of ${\cal I}(t)$. As long as expectation values with respect to the initial state $\ket{\Psi(0)}$ can be specified efficiently, as it is typically the case in problems of state preparation, adiabatic quantum computation or quantum approximate optimization algorithms \cite{plesch2011quantum, albash2018adiabatic, farhi2014quantum},
the state fidelity $F$ can be bounded efficiently by Eq.~\eqref{bound_infidelity}.

\subsubsection{Desired unitary}
\label{subsec:desiredunitary}

Similar to the problem of state preparation, the task of realizing a specific unitary -- the central problem in quantum simulation~\cite{RevModPhys.86.153} --  can also be addressed using the present approach, though there are some key differences.
In the above problem of state preparation, there is not a unique choice of control target ${\cal I}_T$ for a given initial condition ${\cal I}(0)$.
This is different for a control problem with the goal to realize a given unitary ${\cal U}_T$.
In this case, the control target corresponding to an initial condition ${\cal I}(0)$ necessarily needs to read ${\cal I}_T={\cal U}_T{\cal I}(0){\cal U}_T^\dagger$.

If there is a unitary $V$ that commutes with ${\cal I}_T$, {\it i.e.} $[V,{\cal I}_T]=0$,
then also the relation ${\cal I}_T=V{\cal I}_TV^\dagger=V{\cal U}_T{\cal I}(0){\cal U}_T^\dagger V^\dagger$ holds. Reaching the target ${\cal I}_T$ does thus not necessarily verify that the propagator ${\cal U}_T$ is realized. Ensuring that one operator ${\cal I}(t)$ with an initial condition ${\cal I}(0)$ evolves towards the target ${\cal I}_T={\cal U}_T{\cal I}(0){\cal U}_T^\dagger$ that is consistent with the unitary ${\cal U}_T$, therefore, does not necessarily guarantee that the unitary ${\cal U}_T$ is indeed realized.
Similarly to the definition of a faithful objective function in terms of a set of initial and final states, also the present approach requires a set of initial conditions ${\cal I}_j(0)$ and corresponding targets ${\cal I}_{Tj}={\cal U}_T{\cal I}_j(0){\cal U}_T^\dagger$ for the definition of an objective function that is minimized exactly for the desired unitary ${\cal U}_T$. Unlike standard methods for gate characterization, which often require an exponentially large set of initial and final states~\cite{PhysRevA.71.062310,PhysRevResearch.3.033031}, this operator-based approach typically requires only a small number of initial conditions~\cite{PhysRevA.88.042309}.

A possible choice of a faithful objective function is given by
\be
{\cal J}_{{\cal U}_T}= 1 - \frac{1}{N_c}\sum_{j=1}^{N_c} \frac{\mbox{tr}({\cal I}_j(T)\, {\cal U}_T{\cal I}_j(0){\cal U}_T^\dagger)}{\mbox{tr}(({\cal I}_j(0))^2)} 
\label{eq:fidgate}
\ee
with a sufficiently large number $N_c$ of initial conditions $\mathcal I_j(0)$. The initial conditions $\mathcal I_j(0)$ can, for example, always be chosen as a minimal set of operators from $\{ \mathfrak a_j\}$ such that there is no operator in $SU(2^n)$ that commutes with all the transformed operators $\mathcal U_T \mathcal I_j(0) \mathcal U_T^{\dagger}$.  

Since each summand in Eq.~\eqref{eq:fidgate} can be evaluated analogously to Eq.~\eqref{eq:J},
the objective function ${\cal J}_{{\cal U}_T}$ can indeed be evaluated with polynomial effort. 

\subsubsection{Numerical optimization}
\label{Sec:numerial_opt}

Once a control target ${\cal I}_T$ and an objective function, such as ${\cal J}$ (Eq.~\eqref{eq:J}) or ${\cal J}_{{\cal U}_T}$ (Eq.~\eqref{eq:fidgate}) is specified, one can use any numerical pulse shaping algorithm~\cite{Koch2022} to design a time-dependent Hamiltonian that realizes a dynamics from ${\cal I}(0)$ to ${\cal I}(T)$, at a final time $T$, that is close to ${\cal I}_T$.
Since Eq.~\eqref{dynamics_aj} with an anti-symmetric matrix $K$ induces orthogonal (unitary and real) dynamics, any numerical algorithm that is applicable to the Schr\"odinger equation (such as Krotov~\cite{goerz2019krotov}, GRAPE~\cite{Khaneja2005} or CRAB~\cite{doria2011optimal}) is directly applicable, and techniques to construct {\it e.g.} gradients with respect to control parameters also directly apply to Eq.~\eqref{dynamics_aj} (see  Appendix~\ref{app:num_opt} for more details on the implementation).

Since any such optimization is based on numerically exact integration of Eq.~\eqref{dynamics_aj}, there is no restriction to adiabaticity, and the duration $T$ of the control protocol can be chosen as short as allowed by the physical constraints of the underlying system.

\subsection{A birdseye view on the algorithm}

Putting everything together, the resultant methodology for state-control can be summarized in terms of the following algorithm.

Given a Hamiltonian with tuneable time-dependence (Eq.~\eqref{eq:Hamiltonianexpand}) and a set of operators $\{\mathfrak{a}_j\}$, such that the closed commutator relationships (Eq.~\eqref{eq:structure}) are satisfied,
\begin{itemize}
    \item[-] choose an initial condition ${\cal I}(0)=\sum_ja_j(0)\mathfrak{a}_j$, with a non-degenerate ground state. 
    \item[-] find a control target ${\cal I}_T$ that implicitly specifies the target state, and that needs to be reachable from initial condition ${\cal I}(0)$ with the dynamics induced by the system Hamiltonian.
    \item[-] design an optimized time-dependence of the system Hamiltonian (Eq.~\eqref{eq:Hamiltonianexpand}), with the dynamics described by Eq.~\eqref{dynamics_aj}, to minimize an objective function
    formulated in terms of the expansion coefficients of the control target ${\cal I}_T$ in the set of operators $\{\mathfrak{a}_j\}$.
    The design of this time-dependence can be pursued with any pulse-shaping algorithm that applies to linear differential equations.
\end{itemize}
The resultant time-dependent Hamiltonian is such that the system, if initialized in the ground state of ${\cal I}(0)$, evolves towards the ground state of ${\cal I}_T$.

The algorithm for control of unitary dynamics is largely equivalent, but it requires more that one initial condition, and the target for each initial condition is defined uniquely through the desired unitary.

\section{State preparation in a driven spin chain}
\label{sec:statetransfer}

\begin{figure}[t]
\centering
\includegraphics[width=\columnwidth]{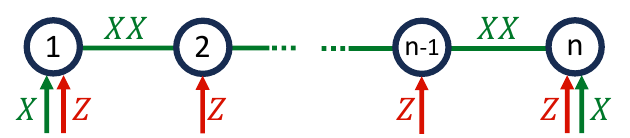}
\caption{Linear spin chain of $n$ qubits with nearest-neighbor $XX$ interactions.
Single-qubit $Z$-driving and $X$-driving on the end qubits enables efficient preparation of various quantum states, including GHZ and cluster states, and the realization of effective dynamics with multi-qubit interactions.}
\label{fig:ising_figure}
\end{figure}

The approach outlined in Sec.~\ref{sec:invariantcontrol} above is applicable to any set of operators  satisfying Eq.~(5). The benefits of this approach depend on the size $d$ of the resulting set $\{ \mathfrak a_j \}$.
There does not seem to be an existing algebraic framework to determine the system Hamiltonians with an associated set of operators $\{ \mathfrak a_j \}$ that is significantly smaller than the size of the system Hilbert space, but the existence of such Hamiltonians can be demonstrated with explicit examples.

Sec.~\ref{sec:scaling} provides a list of exemplary spin systems and sets of operators with a size that is polynomial in the number of spins. Since such a list of examples is necessarily a bit technical, the following discussion and the subsequent section Sec.~\ref{sec:realization_propagators}  will illustrate the application of the general control framework outlined in Section~\ref{sec:invariantcontrol} with the specific case of a linear spin chain with nearest-neighbor interactions, as depicted in Figure~\ref{fig:ising_figure}.

In such a configuration, the system dynamics is induced by the system Hamiltonian
\be
\label{eq:ham}
H(t)=g\sum_{j=1}^{n-1}X_jX_{j+1}+
\sum_{j=1}^{n}f_j(t)Z_j+ \hspace{-0.2cm}
\sum_{j\in\{1,n\}} \hspace{-0.2cm} w_j(t)X_j \ ,
\ee
which consists of a constant interaction term and single-qubit control terms.
The constant term contains nearest-neighbor interactions in terms of the Pauli $X$ operators,
while the single-qubit control terms, with tuneable time-dependent functions $f_j(t)$ and $w_j(t)$, contain the Pauli $Z$ operators and the Pauli $X$ operators on the end-spins of the chain (see Fig.~\ref{fig:ising_figure}).

Without the single-qubit $X$-terms, the Hamiltonian would correspond to the regular Ising model.
The parity conservation of the Ising model imposes limits on the achievable control, and the Ising model admits a treatment in terms of free Fermions~\cite{mbeng2020quantum} that helps to overcome the exponential effort required to simulate the system dynamics.

The Hamiltonian $H(t)$ in Eq.~\eqref{eq:ham}, on the other hand, has no conserved quantities for $g\neq 0$ and general time-dependent functions $f_j(t)$ and $w_j(t)$;
{\it i.e.}, there is no operator $A$ (apart from multiples of the identity) that commutes with all operators $Z_j$, $X_1$, $X_n$ and $\sum_{j=1}^{n-1}X_jX_{j+1}$ as discussed in more detail in Appendix~\ref{sec:no-symmetries}.

The Lie algebra generated by nested commutators of the individual terms in the Hamiltonian (Eq.~\eqref{eq:ham}) contains $d=2n^2+3n+1$ terms as discussed in Sec.~\ref{sec:algebrachain}, and they are given by
\begin{subequations}
\be
Z_{i}\ ,
\label{eq:transLieZ1}
\ee
with
$i \in [1,n]$,
\begin{flalign}
X_{j}& Z_{jk}X_{j+k+1}\ ,
\label{eq:transLie2nd}\\
Y_{j}& Z_{jk}Y_{j+k+1}\ , \\
X_{j}& Z_{jk}Y_{j+k+1}\ ,\\
Y_{j}& Z_{jk}X_{j+k+1}\ ,
\end{flalign}
with
$j\in [1, n-k-1]$, $k \in [0, n-2]$,
and the short-hand notation $Z_{jk}=\prod_{i=1}^{k}Z_{i+j}$,
\begin{flalign}
Z_{0j}&X_{j+1}\ ,\\
Z_{0j}&Y_{j+1}\ ,\\
&X_{n-j} Z_{(n-j)j}\ ,\\
&Y_{n-j} Z_{(n-j)j}
\label{eq:transLie2ndlast}\ ,
\end{flalign}
with $j \in [0, n-1]$, and finally
\be
\prod_{i=1}^{n}Z_{i}\ .
\label{eq:transLieZn}
\ee
\label{eq:transLie}
\end{subequations}

While the system Hamiltonian (Eq.~\eqref{eq:ham}) contains only the terms $Z_j$, $X_jX_{j+1}$, $X_1$ and $X_n$, the operator ${\cal I}(t)$ needs to be expanded in terms of the full set of Eq.~\eqref{eq:transLie} so that Eq.~\eqref{eq:structure} is satisfied and the Ansatz in Eq.~\eqref{eq:Iexpand} is sufficiently general to allow for integration of Eq.~\eqref{eq:eomi}.

Since Eq.~\eqref{eq:transLie} contains a quadratic number of terms, the present framework reduces a control problem in an exponentially large Hilbert space to a problem in a vector space of quadratic growth.
Thanks to this quadratic scaling, it is possible to design optimal control protocols for spin chains of lengths that are inaccessible for approaches based on quantum states.

The following examples are motivated by typical problems of quantum simulation and state preparation or stabilisation.
They exploit the fact that the Lie algebra given in Eq.~\eqref{eq:transLie} contains three-body interactions such as $X_jZ_{j+1}X_{j+2}$ or even $n$-body interactions that cannot be realised by static means. The goal of the following two sections Sec.~\ref{sec:targethamiltonians}
and \ref{sec:propagators_objective} will thus be to find time dependent functions $f_j(t)$ and $w_j(t)$ in the system Hamiltonian $H(t)$ (Eq.~\eqref{eq:ham}) that ensure that the system dynamics with only the two-body interaction in Eq.~\eqref{eq:ham} is helpful for preparation of states or realization of propagators that are characterized by three-body interactions or interactions between more than three spins.

\subsection{Target Hamiltonians}
\label{sec:targethamiltonians}

Following the discussion of Sec.~\ref{subsec:diabatictransitions}, one can seek to find time-dependent functions $f_j(t)$ and $w_j(t)$ in the system Hamiltonian $H(t)$ (Eq.~\eqref{eq:ham}), such that the system evolves from an initial state to a desired final state. Typical choices for an initial state are given by separable states that can be specified efficiently, but final states of interest are often highly entangled states.
The following discussion will thus avoid the explicit specification of a state vector, but both the initial state and the final state to be realized will be specified in terms of a Hamiltonian to which they are the ground state. 

A natural choice of non-interacting Hamiltonian based on Eq.~\eqref{eq:ham} is given by the initial conditions $f_j(0)=f_k(0)$ for all $j,k$,
$f_j(0)\gg g$ and $w_j(0)=0$ for all $j$, {\it i.e.} non-interacting spins with single-spin-Z Hamiltonian.

Crucially, the Hamiltonians defining the final state to be realized do not need to be of the form of the system Hamiltonian in Eq.~\eqref{eq:ham},
but any Hamiltonian comprised of the operators given in Eqs.~\eqref{eq:transLie} including three-spin interactions or interactions between more than three spins is a viable choice. 

As interacting Hamiltonians defining target states, the subsequent discussion encompasses the Hamiltonian
\be
H_G=-\sum_{j=1}^{n-1} X_j X_{j+1}-\prod_{i=1}^{n}Z_{i}\ ,
\label{eq:Hg}
\ee
whose ground state is a GHZ state, which holds significant importance in the fields of quantum sensing \cite{degen2017quantum}, quantum communication \cite{jin2006three, hillery1999quantum}, and macroscopic quantum mechanics \cite{frowis2011stable}, attracting great experimental interest \cite{omran2019generation,song2019generation}.

Another interesting example is the Hamiltonian
\be
H_C=Z_1X_2+\sum_{j=1}^{n-2} X_jZ_{j+1}X_{j+2}+X_{n-1}Z_n\ ,
\label{eq:Hc}
\ee
whose ground state is  the one-dimensional $n$-qubit cluster state for measurement-based quantum computation~\cite{nielsen2006cluster}.

Since the Hamiltonians $H_G$ and $H_C$ admit an analytic construction of a target invariant, the following discussion includes also the example 
\begin{align}
    H_D=H_C+\sum_{j=1}^{n-1}X_jX_{j+1}
    \label{eq:Hd}
\end{align}
to demonstrate that no analytic solutions are necessary preconditions for the present approach. In particular, this Hamiltonian has a vanishing gap in the limit ${n\to\infty}$ which rules out optimal control techniques based on matrix product states~\cite{doria2011optimal} as practical alternatives.

\subsection{Control targets}

As discussed in Sec.~\ref{subsec:diabatictransitions}, a natural choice for the initial condition ${\cal I}(0)$ is given by
\be
{\cal I}_0=\sum_j Z_j
\label{eq:I0_def}
\ee
for the present problem.

The definition of corresponding control target, however, requires some care, and it needs to be done separately for the three different control problems defined in terms of the three Hamiltonians $H_G, H_C$ and $H_D$.

\begin{figure*}[t]
\centering
\includegraphics[width=2\columnwidth]{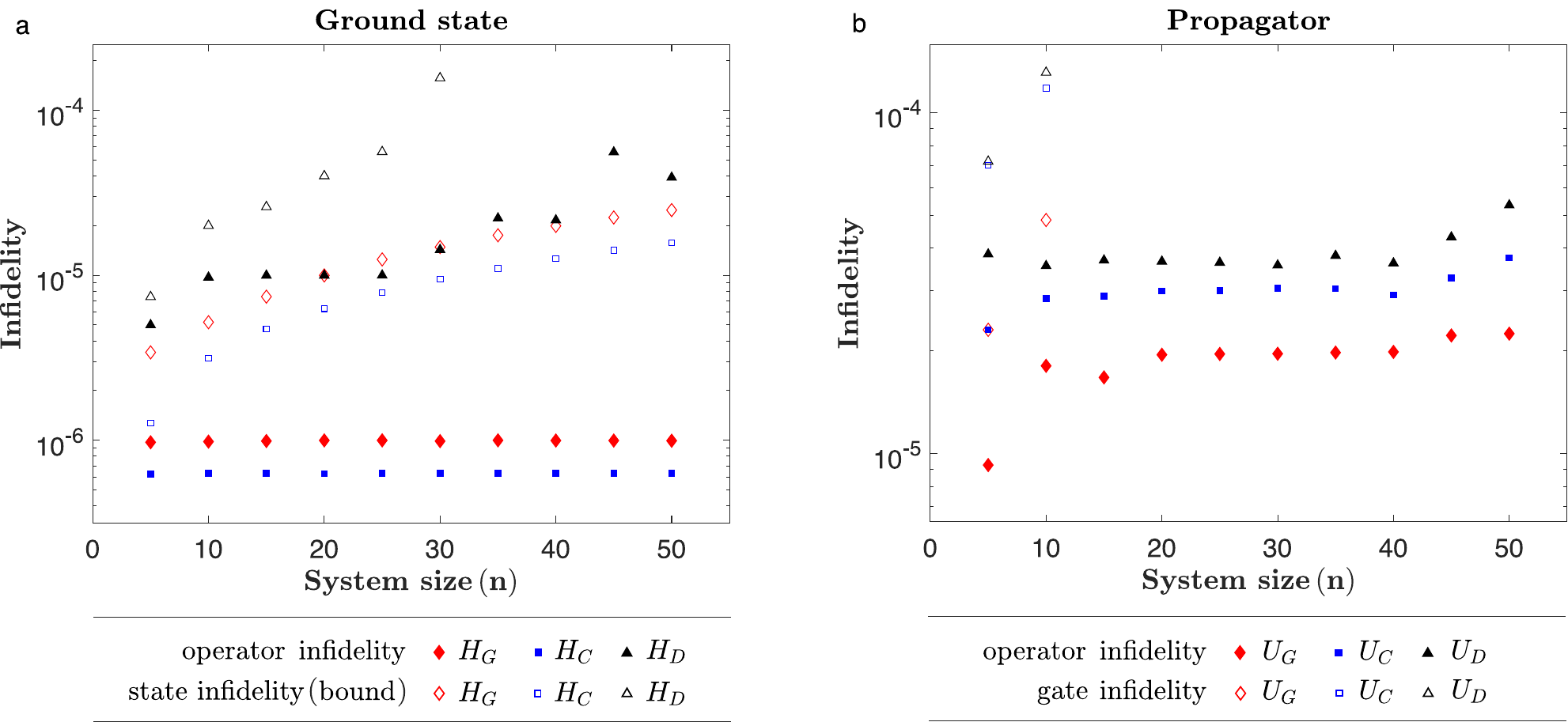}
\caption{Infidelity for dynamics for a spin chain (Eq.~\eqref{eq:ham}) with optimized single-qubit driving, as function of systems size $n$.
Solid shapes depict operator infidelities $\cal J$ (Eq.~\eqref{eq:J})
in log-scale for state transfer (panel a) and operator infidelities ${\cal J}_{U_T}$ (Eq.~\eqref{eq:fidgate})
for quantum gates (panel b).
Empty diamonds and squares in panel a depict bounds (Eq.~\eqref{bound_infidelity})
on state infidelities; empty triangles in panel a represent state infidelities based on matrix-product-state representations and empty shapes in panel b depict gate infidelities.
As result of the optimization, the values of the infidelities $\cal J$ and ${\cal J}_{U_T}$ are largely independent of the system size $n$.
Actual state- and gate-infidelities do grow with increasing system size, but they remain well below what can be achieved given the experimental imperfections of actual quantum devices.}
\label{fig:infid}
\end{figure*}

\subsubsection{Control target for GHZ-state Hamiltonian}
\label{sec:targetGHZ}

Even though it is generally not possible to define the control target ${\cal I}_T$ in terms of the Hamiltonian whose ground state is supposed to be realized, this is possible in the case of $H_G$.
Since $H_G$ can be expressed as
\be
H_G=W_G^\dagger {\cal I}_0 W_G\ ,
\ee
with
\be
W_G=\left( \prod_{k=1}^{n-1} e^{-i \frac{\pi}{4} (X_{k}Y_{k+1})} \right) e^{i \frac{\pi}{4} (X_1 - X_n)}\ ,
\ee
where the factors in the product are in increasing order in the index $k$, the two operators $H_G$ and ${\cal I}_0$ have the same spectrum; and since every factor in $W_G$ is induced by elements of the Lie Algebra (Eq.~\eqref{eq:transLie}), the unitary $W_G$ can indeed be realized with the Hamiltonian $H(t)$ in Eq.~\eqref{eq:ham}.
The control target ${\cal I}_G=H_G$ is thus a valid choice for the task of realising the ground state of $H_G$.

\subsubsection{Control target for cluster-state Hamiltonian}
\label{sec:targetcluster}

The construction of control target ${\cal I}_C$ for the task of realising the ground state of $H_C$ is very similar to the case of $H_G$ above.

$H_C$ can be expressed as
\be
H_C=W_C^{\dagger}\mathcal{I}_0 W_C
\ee
with
\begin{align}
W_C = \left( \prod_{k=1}^{n-1} e^{(-1)^{k} i \frac{\pi}{4}(X_{k}X_{k+1})} \right) e^{i \frac{\pi}{4}(X_1 + X_n)}\ .
\end{align}
Also here, all the terms in $W_C$ are induced by elements of the Lie Algebra (Eq.~\eqref{eq:transLie}).
Since $W_C$ is unitary, the control target ${\cal I}_C=H_C$ is thus a valid choice for the task of realising the ground state of $H_C$ starting with the ground state of ${\cal I}_0$.

\subsubsection{Control target for $H_D$}\label{sec:hd}

While control targets could be defined analytically for the control problem of realizing the ground state of $H_G$ and $H_C$, the case of $H_D$ seems to require a numerical construction. The target $\mathcal I_D=H_D$ is unreachable from $\mathcal I(0)$ with unitary dynamics, so a numerical target $\mathcal I_D$ with the same ground-state as $H_D$ must be constructed.

Propagating ${\cal I}(t)$ according to Eq.~\eqref{eq:eomi} yields operators that are naturally reachable with the available dynamics, and choosing a Hamiltonian $H_{a}(t)$ with an adiabatically slow time-dependence can help to find an operator ${\cal I}_a(T_a)$ whose ground state is also ground state of $H_{a}(T_a)$ at the end of the adiabatic dynamics.

The following example is based on the choice
\be
H_a(t) = \left(1-\frac{t}{T_a}\right) H_I + \frac{t}{T_a} H_D  + \frac{t}{T_a}\left(1-\frac{t}{T_a}\right) H_{B}\ ,
\ee
where the last term with
\be
H_{B} = \sum_j Z_j + \sum_{j=1}^{n-1} X_j X_{j+1} + X_1 + X_n
\ee 
vanishes at the initial and final point in time ($t=0$ and $t=T_a$), but turns some level crossings into avoided crossings, so that adiabatic dynamics does indeed ensure transfer from the ground-state of $H_I$ to the ground-state of $H_D$.

A control target ${\cal I}_D={\cal I}_a(T_a)$ can be constructed numerically with the adiabaticity parameter $T_a$ sufficiently large.
Since any numerical propagation is based on Eq.~\eqref{dynamics_aj}, the numerical effort follows the quadratic scaling of the Lie algebra (Eq.~\eqref{eq:transLie}) and not the exponential scaling of the Hilbert space.

Once the control target is defined, it can be used to construct a time-dependent Hamiltonian that achieves the state transfer in a diabatic fashion.

\subsection{Results -- State transfer}
\label{sec:results_state_transfer}

Fig.~\ref{fig:infid}a depicts the infidelity ${\cal J}$ (Eq.~\eqref{eq:J}) obtained with numerically optimized, time-dependent Hamiltonians as function of the number $n$ of spins in a chain with $n$ ranging from $5$ to $50$.
The different solid symbols correspond to the three different target operators discussed in Sec.~\ref{sec:targethamiltonians}, namely $H_G$ (Eq.~\eqref{eq:Hg}) (red diamonds), $H_C$ (Eq.~\eqref{eq:Hc}) (blue squares) and  $H_D$ (Eq.~\eqref{eq:Hd}) (black triangles).
The optimisations for $H_G$ and $H_C$ are intentionally stopped when infidelities fall below $10^{-6}$. The optimisation for $H_D$, however, requires longer running time and is therefore terminated once a maximum number of iterations, around $10^3$, is reached.

The empty red diamonds and blue squares depict the bound of Eq.~\eqref{bound_infidelity} on the state fidelity for the ground states of $H_G$ and $H_C$. The gap between the infidelities for the operators and the bounds on the state infidelity grows approximately linearly with the number of qubits (for more details, see Appendix~\ref{sec:bound}). For practical implementations, this growing gap  does not seem to pose a relevant limitation, since state infidelities on the order of $10^{-5}$ are below what is currently achievable or verifiable even from few-qubit states~\cite{bharti_noisy_2022}. 

The Hamiltonian $H_D$ is gap-less, {\it i.e.}, the gap between the ground state and the first excited state vanishes in the thermodynamic limit. For finite-size chains, this gap gets progressively smaller with increasing system size, making the bound of Eq.~\eqref{bound_infidelity}
unsuited for $H_D$.
The state fidelity in this case is thus estimated from numerical propagations with matrix product states~\cite{tenpy,Zaletel2015} for qubit numbers $n\leq 30$, as depicted by the empty triangles in Fig.~\ref{fig:infid}a.
Despite the efficient matrix-product representation of the initial and final states, the states at in-between times created by the controlled dynamics for $n\geq 35$ require  bond dimensions exceeding $500$, rendering numerical propagation impractical.
Both operator infidelities (Eq.~\eqref{eq:J}) and state infidelities are larger than for $H_G$ and $H_C$.
This is mostly due to the closing gap of $H_D$ that results in increasing errors in the adiabatic evolution used to obtain the target $\mathcal I_D$. 
The optimization of the control pulses is made consistent with the estimated error of $\mathcal I_D$, {\it i.e.} the optimization algorithm is terminated when infidelities of the order of the accuracy of $\mathcal I_D$ are reached.
For any finite system, this error can be reduced further if needed, and the present accuracy is chosen in accordance with the practical limitations of existing quantum devices~\cite{bharti_noisy_2022}. 

\begin{figure*}[t]
\centering
\includegraphics[width=0.9\textwidth
]{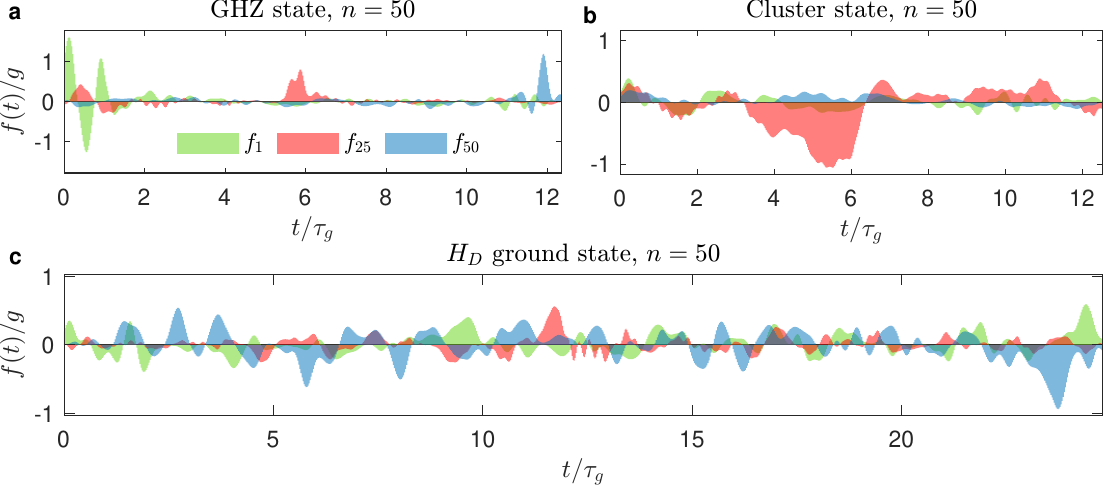}
\caption{Optimised pulses, $f_1$, $f_{25}$ and $f_{50}$, for state transfer in a chain with 50 spins as function of time $t$ in multiples of the interaction time $\tau_g=2\pi/g$.
The operator infidelities obtained with these pulses are $\approx 10^{-5}$ for the GHZ state (a), $\approx 10^{-5}$ for the cluster state (b),  and $\approx 2\times  10^{-5}$ for the ground state of $H_D$ (c).
The duration of controlled dynamics is longer for $H_D$ in (c) than in the other cases.
This is likely due to the difficulty in realising the ground state of a gap-less model.}
\label{fig:pulse_gs}
\end{figure*}

In all three cases shown in cases Fig.~\ref{fig:infid}a, the achieved state fidelities confirm successful optimal control for system sizes far outside the regime accessible to explicit state control. All deviations from perfect, vanishing infidelity are due to finite numerical accuracy.
In Fig.~\ref{fig:infid}, the accuracy is chosen to not exceed the limits imposed by system imperfections of current or foreseeable quantum devices too excessively, but higher accuracies can be reached if technological advances require more accurate solutions.  

Interestingly, because of the constant interaction term in the system Hamiltonian (Eq.~\eqref{eq:ham}), there is a minimal time required for the state transfer, also referred to as quantum speed limit \cite{deffner2017quantum}. In order to obtain infidelities as low as in Fig.~\ref{fig:infid}, the duration of the controlled dynamics needs to increase with the system size, but a modest approximately linear increase $\sim n \tau_g/4$ for $H_G$ and $H_C$, and $\sim n \tau_g/2$ for $H_D$, with the qubit number $n$ and the timescale $\tau_g=2\pi/g$ given by the interaction constant $g$ in Eq.~\eqref{eq:ham}, is sufficient for the ground state transfer. The time-dependence in the system Hamiltonian is parametrized in terms of piecewise constant pulses that are suitable for numerical optimization algorithms. To ensure the linear increase in gate-time while maintaining a constant control duration per time bin across all system sizes, the number of time bins in the pulse must also increase linearly with the system size.
The results in Fig.~\ref{fig:infid} are obtained with pulses of $10 n$ bins, except for $U_G$ with $20n$ bins,
but the exact number can be adjusted depending on the desired infidelity.

Despite the challenging goal of realizing eigenstates of Hamiltonians beyond pairwise interactions, the necessary control pulses do not exhibit undesired characteristics like excessively high amplitudes or frequencies.
Even optimizations without penalty for high amplitudes do not yield amplitudes exceeding the coupling strength $g$ by a factor of $2$.
The spectral width obtained with a standard optimization (see Appendix~\ref{sec:optm}) is not excessive, but it is broader than necessary.
It is thus possible to smooth pulses with only a moderate increase in infidelity, and to use these smoothed pulses as initial condition for further optimization.

Fig.~\ref{fig:pulse_gs} depicts pulse shapes obtained in such a fashion for $f_1(t)$, $f_{25}(t)$ and $f_{50}(t)$ in a chain with $n=50$ spins, {\it i.e.} driving patterns for the spins at the edges and center of the chain. In addition to the smooth temporal shape, the highest amplitudes are of the order of $g$, and even this moderately strong driving arises only in short intervals.

For the GHZ state, the optimised pulses hint at a pattern of sequential driving of the spins from left to right along the chain, as the pulses for the spins on the left peak first.
This is shown in Fig.~\ref{fig:pulse_gs}a for the 1st, 25th and 50th spins, and this sequential pattern does indeed occur for the pulses of the other spins in the chain. 
 
The pulse shapes for smaller chains have qualitatively similar shapes, but both the duration and the number of time bins are approximately proportional to the system size. The pulse data for all spins and system sizes ranging from 5 to 50 is provided in Ref.~\cite{zenodo}. 

\section{Realization of propagators}
\label{sec:realization_propagators}

Preparing an eigenstate of a Hamiltonian with specific interactions has applications in computational problems~\cite{PhysRevA.103.032433, hogg2003adiabatic}.
In quantum simulation~\cite{RevModPhys.86.153}, on the other hand, it is of interest to explore the dynamics induced by a given Hamiltonian. Such applications require -- in addition to an initial state preparation -- the realization of propagators.
Propagators characterized by Hamiltonians beyond pairwise interactions, such as $H_G$, $H_C$, $H_D$ discussed above in Sec.~\ref{sec:targethamiltonians} cannot be realized by static means, and the present approach can be used to find time-dependent functions $f_j(t)$, $w_1(t)$ and $w_n(t)$ in the system Hamiltonian $H(t)$ (Eq.~\eqref{eq:ham}), such that the resultant propagator is of the form $\exp(-iH_TT)$ with a Hamiltonian including more-than-pairwise interactions at a given time $T$. 

The realization of such effective Hamiltonians is commonly pursued based on Floquet theory~\cite{PhysRevX.4.031027} and the present approach allows us to go beyond the common perturbative treatment that is typically required~\cite{Eckardt_2015}. 

The realization of desired propagators
is exemplified here in more detail for $\exp(-i\pi/8\,  H_G)$, $\exp(-i\pi/8\,  H_C)$ and $\exp(-i\pi/8\,  H_D)$ given by the three Hamiltonians $H_G$ (Eq.~\eqref{eq:Hg}), $H_C$ (Eq.~\eqref{eq:Hc}) and  $H_D$ (Eq.~\eqref{eq:Hd}),
and system dynamics in terms of the Hamiltonian $H(t)$ in Eq.~\eqref{eq:ham}.

\subsection{Objective function}
\label{sec:propagators_objective}

The crucial question that will ultimately determine the required numerical effort is how many (and which) initial conditions ${\cal I}_j(0)$ for the infidelity ${\cal J}_{U_T}$ in Eq.~\eqref{eq:fidgate} need to be considered in order to obtain a faithful objective function.

With the initial conditions

\begin{subequations}
\be
{\cal I}_j(0)=Z_j\ ,
\ee
for $j\in[1,n]$,
\be
{\cal I}_{j+n}(0)=X_jX_{j+1}\ ,
\ee
for $j\in[1,n-1]$, and
\be
{\cal I}_{2n}(0)=X_1+X_{n}\ ,
\ee
\label{eq:fullinitcond}
\end{subequations}
the infidelity ${\cal J}_{U_T}$ in Eq.~\eqref{eq:fidgate} vanishes exactly for the target gate $U_T$, for any target that is reachable with $H(t)$ in Eq.~\eqref{eq:ham}.
That is to say that the relation
$U\mathcal{I}_j(0)U^\dagger=U_T\mathcal{I}_j(0)U_T^\dagger$ can be simultaneously satisfied for all these choices $\mathcal{I}_j(0)$ only if $U=U_T\exp(i\varphi)$, where $\exp(i\varphi)$ is a global phase factor.

To see that this is indeed the case, it is helpful to notice that the set of relations 
$U\mathcal{I}_j(0)U^\dagger=U_T\mathcal{I}_j(0)U_T^\dagger$ is equivalent to the set of commutation relations
\be
[U_T^\dagger U,\mathcal{I}_j(0)]=0\ .
\ee
As shown in Sec.~\ref{sec:no-symmetries} of the Appendix, the only operators that commute with each of the operators in Eq.~\eqref{eq:fullinitcond} are multiples of the identity, which leaves $U_T^\dagger U=\exp(i\varphi)\mathbbm{1}$ as the only solution.

While this provides a faithful objective function, its use  in an optimisation implies the  evaluation of $2n$ summands in Eq.~\eqref{eq:fidgate} at any step in the optimisation.
For the sake of efficiency, it can be preferable to use the objective functional ${\cal J}_{{\cal U}_T}$
 with only five initial conditions
\begin{subequations}
\bqa
{\cal I}_1(0)&=&\sum_kZ_{2k+1}\ ,\\
{\cal I}_2(0)&=&\sum_{k}Z_{2k}\ ,\\
{\cal I}_3(0)&=&\sum_{k}X_{2k+1}X_{2k+2}\ ,\\
{\cal I}_4(0)&=&\sum_{k}X_{2k}X_{2k+1}\ ,\\
{\cal I}_5(0)&=&X_1+X_n\ .
\eqa
\label{eq:redinitcond}
\end{subequations}
\begin{figure*}[t]
\centering
\includegraphics[width=0.9\textwidth
]{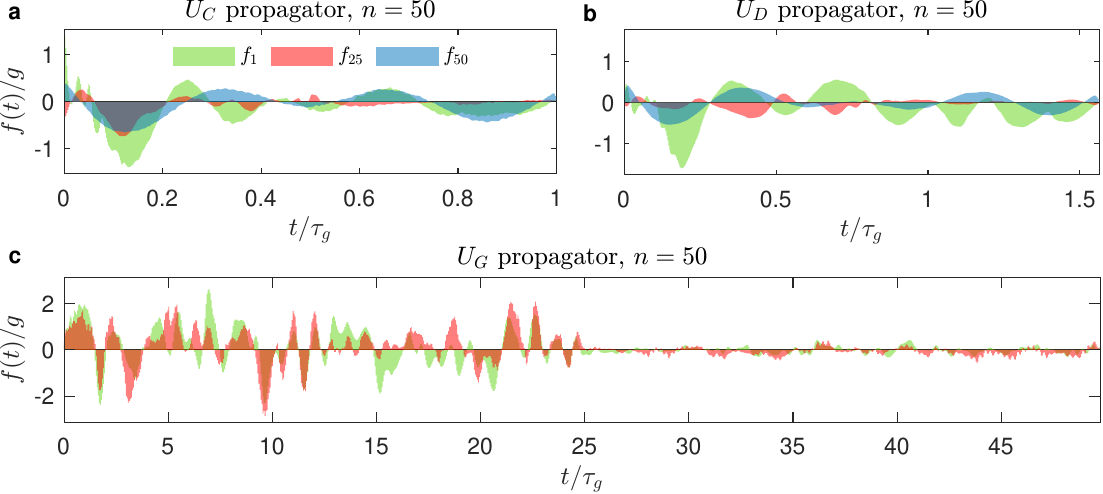}
\caption{Optimised pulses, $f_1$, $f_{25}$ and $f_{50}$, for the realization of propagators in a chain with 50 spins as function of time $t$ in unit of $\tau_g$.
The resultant operator infidelities are $\approx 4 \times 10^{-5}$ for $U_C$ (a), $\approx 9\times 10^{-5}$ for $U_D$ (b), and $\approx 10^{-4}$ for $U_G$  (c). For $U_G$ $f_{50}=f_1$ and hence is not shown.
The pulses for $U_G$ in (f) are longer due to the difficulty in quantum-simulating an $n$-body interaction with a pairwise controlled Hamiltonian.}
\label{fig:pulse_prop}
\end{figure*}

Even though the optimal value can be obtained for propagators different than $U_T$,
optimizations do not seem to converge to such a solution in practice.
Since verification requires only a single propagation, it is numerically advantageous to use this objective function (with Eqs.~\eqref{eq:redinitcond} as initial conditions) for the optimisation, and eventually use the faithful objective function (with Eqs.~\eqref{eq:fullinitcond} as initial conditions) only for verification of the final solution.

\subsection{Results -- Propagators}
\label{sec:propagators}

Fig.~\ref{fig:infid}b depicts the infidelities ${\cal J}_{{\cal U}_T}$, defined in Eq.~\eqref{eq:fidgate} together with Eq.~\eqref{eq:fullinitcond}, as function of system size $n$. The solid red diamonds, blue squares, and black triangles correspond to the target unitaries $U_G=\exp(-i\pi/8\,  H_G)$,
$U_C=\exp(-i\pi/8\,  H_C)$ and
$U_D=\exp(-i\pi/8\,  H_D)$, respectively.
This is based on optimized piecewise constant pulses with a duration of $\sim n \tau_g$ and $20 n$ bins for $U_G$, and $\sim \tau_g$ and $10 n$ bins for $U_C$ and $U_D$. 

The empty shapes in the figure show actual gate infidelities, $1-\left\vert\text{tr}\left(U^{\dagger}(T)U_T\right)\right\vert^2/2^{2n}$, obtained with numerically exact integration of the Schr\"odinger equation.
Since this type of verification is limited by the exponential scaling of the Hilbert space, actual gate infidelities are shown for up to $n=10$ only. 
Within what is numerically achievable, the gate infidelities do confirm the success of the control protocols designed with the present framework,
and similarly to the task of state preparation, the realisation of desired propagators is achieved with infidelities far below of what could be experimentally resolved with existing technology~\cite{horowitz2019quantum, gyongyosi2019survey}.

Also, the duration of controlled dynamics required for the realization of a desired propagator has a fundamental limit given by the timescale $\tau_g=2\pi/g$.
The numerically observed minimal time required to implement the unitaries induced by $H_C$ and $H_D$ is independent of the system size, but the minimal time required to implement the unitary induced by $H_G$ scales linearly in the number of qubits. 
Given that $H_C$ and $H_D$ contain at most three-body interactions, whereas $H_G$ contains an $n$-body interaction, {\it i.e.} an interaction whose type changes with increasing system size $n$,
the numerically observed dependence of minimal durations of controlled dynamics seem consistent with the underlying problem, which generates some confidence that the obtained solutions are close to the actual global solution of the control problem. 

Fig.~\ref{fig:pulse_prop} depicts optimized driving patterns $f_1$, $f_{25}$ and $f_{50}$ obtained with the same procedure of smoothing and reoptimization as in Fig.~\ref{fig:pulse_gs}. Given the fundamental difficulty of quantum simulating the n-body interaction $\prod_{i=1}^nZ_i$ in $U_G$ with a system Hamiltonian including only pairwise interactions, the required duration of the controlled dynamics is substantially longer than in the cases of $U_C$ and $U_D$.
Quite surprisingly, the second half of the time-period shown in panel c is essentially uncontrolled dynamics.
Nevertheless, the full duration of $50\tau_g$ seems to be necessary in order to reach high fidelities, but the controlled dynamics during the first half seems to drive the system to a state that evolves mostly under the influence of the $XX$ interactions towards the target state in the interval $[25\tau_g,50\tau_g]$. The pulse data for smaller system sizes are provided in Ref.~\cite{zenodo}. 

\section{MORE SETS OF OPERATORS WITH POLYNOMIAL SCALING}
\label{sec:scaling}

While the discussion in Secs.~\ref{sec:statetransfer} and \ref{sec:realization_propagators} is meant to exemplify the use of the current approach for actual control problems, a similarly detailed discussion for further examples would be repetitive.
The goal of this section is thus to provide further insight into sets of operators for which the present approach provides an advantage over treatments in terms of state vectors. In the following, a few exemplary algebras with polynomial scaling are provided for different interaction geometries.
The corresponding commutation relations are available at~\cite{zenodo}.

\subsection{Spin chain}
\label{sec:algebrachain}

The Lie algebra specified in Eq.~\eqref{eq:transLie} is generated by the terms
\begin{subequations}
\be
\mathfrak h_j=Z_{j}\ ,
\label{eq:hchainZ}
\ee
with $j\in[1,n]$,
\be
\mathfrak h_{j+n}=X_{j}X_{j+1}\ ,
\label{eq:hchainX}
\ee
with $j\in[1,n-1]$, and
\bqa
\mathfrak h_{2n}&=&X_1\ ,\label{eq:hchainend1}\\
\mathfrak h_{2n+1}&=&X_n\ .\label{eq:hchainend2}
\eqa
\label{eq:Hchain}
\end{subequations}

Commutators between terms in Eq.~\eqref{eq:hchainZ} and terms in Eq.~\eqref{eq:hchainX} yield operators of the form $X_jY_{j+1}$ and $Y_jX_{j+1}$, and commutators between those and terms in Eq.~\eqref{eq:hchainZ} yields operators of the form $Y_jY_{j+1}$.
The Lie algebra thus contain the terms
\bqa
X_{j}X_{j+1}\ ,\ X_{j}Y_{j+1}\ ,\ Y_{j}X_{j+1}\ ,\ Y_{j}Y_{j+1}\ .\label{eq:hchaininteractions}
\eqa
Commutators between such terms yield operators of the form $A_jZ_{j+1}B_{j+2}$ (with $A$ and $B$ $\in\{X,Y\}$),
and nested commutators with several terms of Eq.~\eqref{eq:hchaininteractions} yield operators of the form
\be
A_jZ_{j+1}\hdots Z_{j+k}B_{j+k+1}\ .
\label{eq:hchainmanydoyinteraction}
\ee
Commutators between these $k+2$--body interactions and the terms $X_1$ and $X_n$ (Eqs.~\eqref{eq:hchainend1} and \eqref{eq:hchainend2}) yields the terms
\begin{subequations}
\begin{flalign}
Z_1\hdots Z_l &A_{l+1}\ ,\label{eq:hchainlastZX}\\
&A_{n-l}Z_{n-l+1}\hdots Z_n\ ,\label{eq:hchainlastXZ}
\end{flalign}
and
\be
Z_1\hdots Z_n\label{eq:hchainlastZZ}\ .
\ee
\end{subequations}
Those are exactly the terms specified in Eq.~\eqref{eq:transLie}, and they do indeed form a closed Lie algebra.

With $n$ terms in Eq.~\eqref{eq:hchainZ}, $2n(n-1)$ terms in Eq.~\eqref{eq:hchainmanydoyinteraction} (that include Eq.~\eqref{eq:hchainX} and Eq.~\eqref{eq:hchaininteractions} as special cases), $2n - 1$ terms from Eqs.~\eqref{eq:hchainlastZX} and \eqref{eq:hchainlastXZ} each,
as well as one term in Eqs.~\eqref{eq:hchainend1}, Eqs.~\eqref{eq:hchainend2} and  \eqref{eq:hchainlastZZ} respectively, the total number of elements in the Lie algebra is $d=2n^2 +3n +1$.

\subsection{Driven spin comb}\label{sec:spin_comb}
\label{sec:algebracomb}

\begin{figure}[t]
\centering
\includegraphics[width=0.9\columnwidth
]{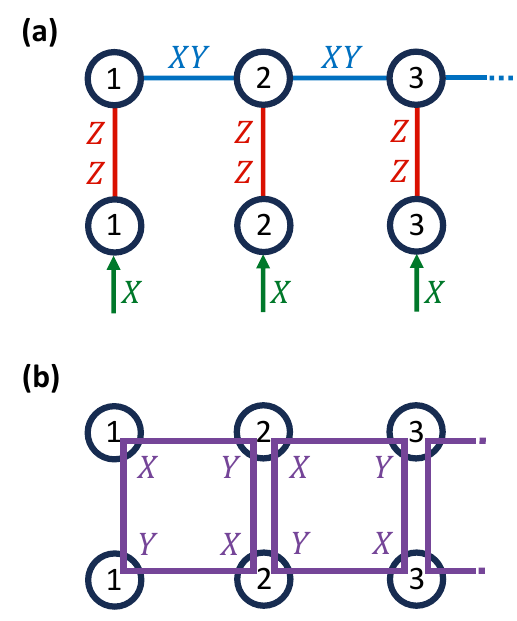}
\caption{Spin comb with nearest-neighbor $XY$ interactions in the top chain and $ZZ$ interactions between spins of different chains (inset (a)).
Single-spin $X$-driving (inset (a)) enables the realization of effective dynamics with four-spin interactions matching the Toric Code Plaquette operators Eq.~\eqref{eq:plaquette} (inset (b)).
}
\label{fig:2D_5}
\end{figure}

An interaction geometry that results in a Lie algebra of quadratic size in the qubit number $n$ that can be used to realize elementary building blocks of topologically protected quantum information processing \cite{stern2013topological} is depicted in Fig.~\ref{fig:2D_5}(a).
It is a comb-structure with a linear chain of $m=n/2$ qubits with nearest-neighbor interactions, and a second chain of another $m$ qubits interacting only with qubits on the first chain. 

Because of this geometry, each Pauli operator carries two indices. The first (lower) index specifies the position of the qubit within its chain, and the second (upper) index identifies the chain itself ($1$ for the top chain and $2$ for the bottom chain).

The elementary Hamiltonian terms $\mathfrak h_i$ (for the definition of a system Hamiltonian in Eq.~\eqref{eq:Hamiltonianexpand}) are given by
\begin{subequations}
\be
\mathfrak h_{j}=Z_{j}^1Z_{j}^2\ ,
\label{eq:hcombZ}
\ee
with $j\in[1,m]$,
\be
\mathfrak h_{j+m}=X_{j}^2\ ,
\label{eq:hcombX}
\ee
with $j\in[1,m]$, and
\be
\mathfrak h_{j+2m}=X_{j}^1Y_{j+1}^1\ ,
\label{eq:hcombXY}
\ee
with $j\in[1,m-1]$.
\label{eq:Hcomb}
\end{subequations}

In order to understand the Lie algebra generated by Eqs.~\eqref{eq:Hcomb}, and, in particular its scaling with $m$, it is instructive to start with the Lie algebra generated by the terms in Eq.~\eqref{eq:hcombXY}  only.
The terms in Eq.~\eqref{eq:hcombXY}  form a subset of the terms in Eq.~\eqref{eq:hchaininteractions}, and similarly to the discussion in Sec.~\ref{sec:algebrachain} the resultant Lie algebra is comprised of the terms
\bqa
&&X_j^1Z_{jk}Y_{j+k+1}^1\ ,\label{eq:hcombXZX}
\eqa

with $j+k+1\le m$ and the short-hand notation $Z_{jk}=\prod_{i=1}^{k}Z_{i+j}^1$.

Commutation between the terms in Eq.~\eqref{eq:hcombXZX} and the terms in Eq.~\eqref{eq:hcombZ}
yields the terms

\begin{subequations}
\bqa
&&Y_j^1Z_{jk}Y_{j+k+1}^1 \ Z_{j}^2\ ,\label{hcombstaticalgebraZstart}\\
&&X_j^1Z_{jk}X_{j+k+1}^1 \ Z_{j+k+1}^2\ ,\label{hcombstaticalgebraZend}\\
&&Y_j^1Z_{jk}X_{j+k+1}^1 \ Z_{j}^2Z_{j+k+1}^2\ .
\label{hcombstaticalgebraZZ}
\eqa
\label{eq:hcombstaticalgebra}
\end{subequations}

With $m(m-1)/2$ terms in Eq.~\eqref{eq:hcombXZX}
(that includes the elements in Eq.~\eqref{eq:hcombXY}),
$3$ terms in Eq.~\eqref{eq:hcombstaticalgebra}, and $m$ elements in Eq.~\eqref{eq:hcombZ}, the Lie algebra generated by the terms in Eq.~\eqref{eq:hcombZ} and Eq.~\eqref{eq:hcombXY} thus contain $2m^2-m$  terms.

Commutation with the terms in Eq.~\eqref{eq:hcombX} yields terms of similar structure to the terms in Eq.~\eqref{eq:hcombstaticalgebra},
but with operators $Z_{j}^2$ and/or $Z_{j+k+1}^2$ replaced by the corresponding operator of $X$ or $Y$ type.
The number of operators of the type in Eqs.~\eqref{hcombstaticalgebraZstart} and \eqref{hcombstaticalgebraZend} is thus increased by a factor of $3$, and 
the number of operators of the type in Eq.~\eqref{hcombstaticalgebraZZ} is increased by a factor of $9$.
The Lie algebra generated by all the terms in Eq.~\eqref{eq:Hcomb} thus has $3m(3m-1)/2$ terms.

Within this Lie algebra are terms of the form
\bqa
Q_j=X_j^1Y_{j+1}^1 \ Y_{j}^2X_{j+1}^2\ ,
\label{eq:plaquette}
\eqa
corresponding to a four-body interaction on four spins forming a square as depicted in Fig.~\ref{fig:2D_5}(b).
The operators $Q_j$ and $Q_{j+1}$ act on two different quartets of spins with an overlap on two spins (with index $j+1$).
The factor of $Q_j$ on these spins is given by $Y_{j+1}^1X_{j+1}^2$, and
the factor of $Q_{j+1}$ on these spins is given by $X_{j+1}^1Y_{j+1}^2$. All the operators $Q_j$ are thus mutually commuting, even though their single-spin factors on overlapping spins are not mutually commuting.
The present example is thus suitable for the realization of a ribbon of plaquette operators of the toric code~\cite{Kitaev_2003}.

\subsection{Hexagonal spin ladder}\label{sec:honeycomb}

\begin{figure}[t]
\centering
\includegraphics[width=0.99\columnwidth]{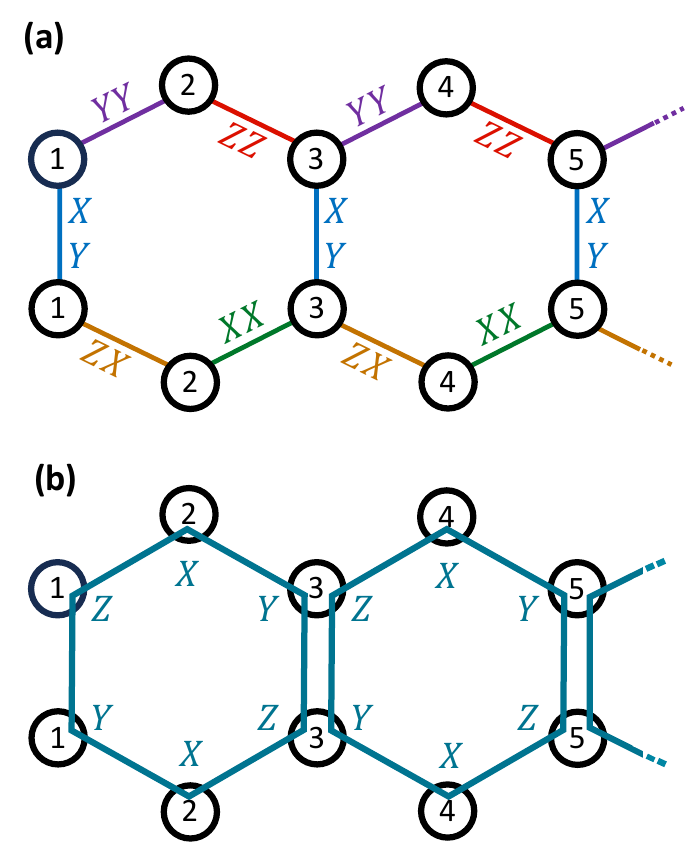}
\caption{Spin-ladder with hexagonal structures.
Each spin interacts with its nearest neighbours, but in contrast to the Kitaev honeycomb model that includes only mutually non-commuting interactions for any spin, there is one spin (at the bottom of each hexagon) that has two commuting interactions (inset (a)).
This interaction geometry results in a Lie algebra of quadratic size with elements including the plaquette operators (Eq.~\eqref{eq:plaquette_op}, inset (b)) that are crucial in the field of topological stabilizer codes.
}
\label{fig:2D_4}
\end{figure}

Fig.~\ref{fig:2D_4} depicts a spin-ladder with hexagonal structures, similar the Kitaev honeycomb model~\cite{Kitaev_2003},
but with the two interactions of the spin at the bottom of each hexagon commuting with each other.
This interaction geometry results in a Lie-algebra of dimension $(25n^2-40n+12)/32$ with $n=4n_h+2$ spins, where $n_h$ is the number of hexagons. In contrast, the interaction geometry of the original Kitaev honeycomb model scales exponentially ($2^{n_h}n\left(n-1\right)/2$).

The individual operators $\mathfrak h_j$ entering the system Hamiltonian read
\begin{subequations}
\bqa
\mathfrak h_j&=& Y_{2j-1}^1 Y_{2j}^1 \ ,
\label{eq:Hhoney_topYY}\\
\mathfrak h_{j+(m-1)/2}&=& Z_{2j}^1 Z_{2j+1}^1 \ ,
\label{eq:Hhoney_topZZ}\\
\mathfrak h_{j+m-1}&=& X_{2j}^2 X_{2j+1}^2 \ ,
\label{eq:Hhoney_bottomXX}\\
\mathfrak h_{j+(3m-3)/2}&=& Z_{2j-1}^2 X_{2j}^2 \ ,
\label{eq:Hhoney_bottomZZ}\\
\mathfrak h_{l+2m-2}&=& X_{2l-1}^1 Y_{2l-1}^2 \ ,
\label{eq:Hhoney_inter}
\eqa
\label{eq:Hhoney_mod}
\end{subequations}
with $m=n/2$, 
$j\in [1, (m-1)/2]$, $l\in [1, (m+1)/2]$.
The resultant Lie algebra can be constructed systematically similarly to the discussion in Sec.~\ref{sec:algebrachain};
because of the more intricate interaction geometry the exact form of the individual elements is a bit bulkier, and it is deferred to Appendix~\ref{lie_hexagonal}.

Crucially, the Lie algebra contains the $6$-body plaquette operators
\begin{align}
     P_j &= Z_{2j-1}^1X_{2j}^1Y_{2j+1}^1Y_{2j-1}^2X_{2j}^2Z_{2j+1}^2 \ , 
\label{eq:plaquette_op}
\end{align}
as depicted schematically in Fig.~\ref{fig:2D_4}b.
These operators, that commute with the Kitaev honeycomb Hamiltonian~\cite{Kitaev_2003}, form a central part in the realization of error-correcting surface codes on hexagonal lattices~\cite{paetznick2023performance,gidney2021fault}. In particular, plaquette operators together with two-body operators like $X_{2l-1}^1Y_{2l-1}^2$, can form the basis for stabilizer operators, as exemplified by the $XYZ^2$ topological stabilizer code~\cite{srivastava2022xyz}.
The present framework can thus be used to aid the realization of quantum error correction.

\section{Discussion and Outlook}

The techniques discussed in this paper integrate into and form a part of the spectrum of quantum optimal control techniques.
They are inspired by control techniques that are based on analytic forms ${\cal I}_a(t)$ of dynamical quantum invariants.
For the explicit system Hamiltonian $H(t)$ (Eq.~\eqref{eq:ham}) discussed in this paper, such an analytic form ${\cal I}_a(t)$ would need to be such that a Hamiltonian that satisfies Eq.~\eqref{eq:eomi} together with ${\cal I}_a(t)$ contains only the terms $Z_j$ ($j\in[1,n]$), $X_1$, $X_n$ and the interaction term $\sum_{j=1}^nX_jX_{j+1}$, but none of the additional terms listed in Eq.~\eqref{eq:transLie}.
Finding such an analytic expression seems to be an insurmountable task, and the main goal of this paper is to avoid the dependence on such analytic expressions. 

Notable existing approaches with goals similar to those pursued in this paper include  counterdiabatic driving~\cite{claeys2019floquet,schindler2023counterdiabatic, yao2021reinforcement,vcepaite2023counterdiabatic}, which requires additional terms in the system Hamiltonian to ensure that system dynamics follow adiabatic evolution, even beyond the regime where the adiabatic approximation holds.
For the system Hamiltonian $H(t)$ (Eq.~\eqref{eq:ham}), counterdiabatic terms would typically include terms listed in Eq.~\eqref{eq:transLie}, but that are not contained in the system Hamiltonian in Eq.~\eqref{eq:ham}.
Such terms can typically not be realized exactly, but they can be approximated as the result of higher-order processes in a time-dependent Hamiltonian~\cite{claeys2019floquet}. The present approach aims at avoiding such approximations, and the existence of a closed commutation relation Eq.~\eqref{eq:structure} enables doing this in a numerically exact fashion.

The realization of propagators defined by Hamiltonians that cannot be realized in practice by static means discussed in Sec.~\ref{sec:realization_propagators} is commonly pursued in the context of Floquet engineering~\cite{verdeny2014optimal, verdeny2016quasi}.
While Floquet theory guarantees the existence of such propagators at periodically occurring points in time for periodically time-dependent Hamiltonians, it does not provide us with the ability to construct such propagators exactly.
Existing approaches are thus typically based on perturbative constructions.
Also in this context, the present approach can be helpful to overcome the restriction to perturbative regimes and to find numerically exact solutions.

The construction of explicit control solutions in this work makes use of the popular algorithm for computing the gradients with respect to control parameters utilized in GRAPE \cite{Khaneja2005,deFouquieres2011}, but the structural equivalence between the Schr\"odinger equation and Eq.~\eqref{dynamics_aj} implies that the full range of optimal quantum control algorithms is applicable to the present approach.
For example, more recent methods to construct gradients~\cite{dalgaard2022fast}, alternative parametrizations of the control functions (\textit{e.g.} in terms of Gaussian pulses~\cite{doria2011optimal}), or formulations that do not require back-propagation during optimization ~\cite{machnes2018tunable} could be employed to achieve smoother solutions and potentially enhance convergence of optimal control algorithms. 

While the exponential scaling of composite quantum systems poses fundamental constraints that cannot be overcome in general, the fact that there are sets of operators with favourable scaling does allow us to overcome them in specific cases.
With control over interaction geometries in various platforms such as superconducting qubits~\cite{wendin2017quantum}, trapped ions and Rydberg atoms~\cite{ebadi2021quantum} the present approach can help to fully exploit the qubit count in current devices (ranging up to $256$~\cite{ebadi2021quantum}) without the need of long gate sequences that are conflicting with decoherence and the accumulation of individual gate errors.
While verification of the functionality of a quantum device is a challenging problem~\cite{Carrasco2021}, the description of quantum dynamics in terms of invariants can also be used to make predictions for the targeted dynamics that can be experimentally tested.

With three-body and four-body interactions naturally arising in the dynamics induced by the spin Hamiltonian given in Eq.~\eqref{eq:ham}, the present approach can also support the realisation of error correcting codes~\cite{cory1998experimental, chiaverini2004realization} and topologically protected quantum information processing~\cite{stern2013topological}.
With the advent of highly controllable systems that pose control problems beyond the limitations of classical simulations~\cite{ebadi2021quantum}, there is a growing need for control techniques with good scaling behaviour.
Optimisations with the present framework can help to avoid or support costly experimental optimisations~\cite{omran2019generation, sauvage2020optimal}.
The resulting ability to create quantum states that allow for sensing at the Heisenberg limit makes the present approach a valuable tool for the community's effort towards the development of quantum technological applications.

\subsection*{Acknowledgements}
This work was supported by the U.K. Engineering and Physical Sciences Research Council via the EPSRC Hub in Quantum Computing and Simulation (EP/T001062/1) and the UK Innovate UK (project number 10004857).
We are indebted to stimulating discussions with Leonardo Banchi, Man Hei Cheng and Oliver Reardon-Smith. The optimised pulse and computer code for this work are available without restriction \cite{zenodo}.

M.O.-R. and N.H.L contributed equally to this work.

\appendix

\section{Absence of symmetries in the system Hamiltonian}\label{sec:no-symmetries}

We will show here that there is no symmetry in the Hamiltonian
\begin{align}
H(t)= \sum_{j=1}^n f_j(t) Z_j + g\sum_{j=1}^{n-1} X_j X_{j+1} + \sum_{j\in\{1,n\}}w_j(t)X_j
\label{eq:H}
\end{align}
for $g\neq 0$ and general time-dependent functions $f_j(t)$ and  $w_j(t)$,
{\it i.e.}\begin{quote}
 there is no $n$-qubit operator $A^{n}$ (besides multiples of the identity) that commutes with each of the operators $Z_1,Z_2,\dots, Z_n$, $X_1$, $X_n$ and $\sum_{j=1}^nX_jX_{j+1}$.
\end{quote}

The requirement to commute with each of the operators $Z_1,Z_2,\dots, Z_n$ implies that $A^n$ can be expanded in terms of tensor-products of $\mathbbm{1}$ and $Z$ terms only.

The requirement to also commute with $X_1$ and $X_n$ then implies that $A_n$ is of the form
\be
A^n=\mathbbm{1}\otimes D^n\otimes\mathbbm{1}\label{eq:An}\ ,
\ee
with an $n-2$-qubit operator $D^n$ that can be expanded in terms of tensor-products of $\mathbbm{1}$ and $Z$ terms,
or more explicitly as
\be
A^n=\mathbbm{1}_1\otimes F^n\otimes\mathbbm{1}_{n-1}\otimes\mathbbm{1}_n+\mathbbm{1}_1\otimes G^n\otimes Z_{n-1}\otimes\mathbbm{1}_n\ ,
\ee
with $n-3$-qubit operators $F_n$ and $G_n$ that can be expanded in terms of tensor-products of $\mathbbm{1}$ and $Z$ terms,

In order to address commutativity with
\be
\sum_{j=1}^{n-1}X_jX_{j+1}=H_x^n\ ,
\ee
it is instructive to express $H_x^n$ as
\be
H_x^n=H_x^{n-1}\otimes\mathbbm{1}_n+\mathbbm{1}_{1\hdots n-1}\otimes X_{n-1}\otimes X_n\ .
\ee
The commutator $[A^n,H_x^n]$ thus reads
\bqa
[A^n,H_x^n]&=&
[\mathbbm{1}_1\otimes G^n\otimes Z_{n-1},H_x^{n-1}]\otimes\mathbbm{1}_n\label{eq:aboveonlytermwithX_n}\\
&+&
2i\mathbbm{1}_1\otimes G^n\otimes Y_{n-1}\otimes X_n\label{eq:onlytermwithX_n}\\
&+&
[\mathbbm{1}_1\otimes F^n\otimes\mathbbm{1}_{n-1},
H_x^{n-1}]\otimes\mathbbm{1}_n\ .
\eqa
Since the term in~\eqref{eq:onlytermwithX_n} is the only term with a factor $X_n$, the commutator can vanish only if $G^n$ vanishes.
If this is the case, however, then also the right-hand-side in~\eqref{eq:aboveonlytermwithX_n} vanishes,
such that the commutator reduces to
\be
[A^n,H_x^n]=
[\mathbbm{1}_1\otimes F^n\otimes\mathbbm{1}_{n-1},
H_x^{n-1}]\otimes\mathbbm{1}_n\ ,
\ee
with an operator
\be
A^{n-1}=\mathbbm{1}_1\otimes F^n\otimes\mathbbm{1}_{n-1}\ ,
\ee
that is exactly of the form of Eq.~\eqref{eq:An}, but for $n-1$ instead of $n$ qubits.

The result to be proven thus holds for an $n$-qubit system, provided that it holds for an $n-1$ qubits system,
and since one can readily verify it explicitly for a $2$ qubits system, the proof is completed by induction.

\section{Numerical simulation and optimization}\label{app:num_opt}

This section describes the numerical techniques used to find the optimal control strategies for the problems of state transfer and realization of propagators, as detailed in Secs.~\ref{sec:statetransfer} and \ref{sec:realization_propagators}. The optimal solutions minimize a fidelity measure (defined in Eq.~\eqref{eq:J} for state transfer and in Eq.~\eqref{eq:fidgate} for propagator realization) by finding a suitable time dependence of the 
system Hamiltonian in Eq.~\eqref{eq:ham}. 

The operator is propagated with the equation of motion Eq.~\eqref{dynamics_aj}, where the matrix $K(t)$ is given by the structure constants (Eq.~\eqref{eq:Kmatrix2}). We provide the structure constants for the Lie algebra of Eq.~\ref{eq:transLie} in a computer readable format in Ref.~\cite{zenodo}. One can obtain these structure constants by computing the commutator relation between the basis elements either analytically, or numerically with the help of the Pauli Algebra package in Mathematica \cite{you_pauli_algebra}.

\subsection{Coefficient vectors of the initial and target operators}

In our method, numerical propagation and optimisation are done entirely with the coefficients in a basis expansion of the operators, i.e., the coefficients $a_j$ in the expansion 
\be
{\cal I}=\sum_j a_j \mathfrak{a}_j \ ,
\ee
where $\mathfrak{a}_j$ are the basis elements listed in Eq.~\eqref{eq:transLie}. In this section, we describe how to find these coefficient vectors for the initial and target operators discussed in the paper.

\subsubsection{State preparation}\label{sec:adb}
The initial operator $\mathcal I_0$ (Eq.~\eqref{eq:I0_def}) and the targets $\mathcal I_G\equiv H_G$ (Eq.~\eqref{eq:Hg}) and $\mathcal I_C\equiv H_C$ (Eq.~\eqref{eq:Hc}) are already written as explicit expansions of the basis elements, so the coefficients can be readily read off. 

The control target $\mathcal I_D$ with the same ground state as $H_D$ is obtained by evolving the initial operator ${\cal I}_0$ under the adiabatic Hamiltonian
\begin{align}
&H_a(t)=(1-t/T_a)\sum_j Z_j  +(t/T_a)H_D\nonumber \\
&+(t/T_a)(1-t/T_a)\left(\sum_j Z_j+\sum_{j=1}^{n-1}X_jX_{j+1}+X_1+X_n\right),
\end{align}
where $T_a$ is a sufficiently long duration. In the $d-$dimensional space of the Lie algebra, $H_a(t)$ corresponds to
\begin{align}
&K_a(t)=(1-t/T_a)\sum_j K^z_j+(t/T_a)K_D\nonumber \\
&+(t/T_a)(1-t/T_a)\left(\sum_j K^z_j+\sum_{j=1}^{n-1}K^{xx}_j+K^x_1+K^x_n\right) \ ,
\end{align}
where $K^z_j,K_D, K^{xx}_j$, and $K^x_j$ are the $d\times d$ adjoint representation matrices of $Z_j, H_D, X_jX_{j+1}$ and $X_j$, respectively (see Eq.~\eqref{eq:Kmatrix2}). Starting with the initial coefficient vector $\bm{a}_0$ of $I_0$, one propagates with
\be
\dot{\bm{a}}(t)=K_a(t)\bm{a}(t) \ ,
\ee
for the duration $T_a$ to obtain the coefficient vector $\bm{a}_D$ of $I_D$. 

\subsubsection{Propagators}
Realizing a propagator $U_T$ requires more than a single pair of initial and target operators $\{\mathcal I(0), U_T \mathcal I(0) U_T^{\dagger} \}$. Instead, $2n$ pairs of initial and final operators, $\{\mathcal I_j(0),\mathcal I_{T j}\equiv U_T \mathcal I_j(0) U_T^{\dagger}\}$, are sufficient, with the initial operators $\mathcal I_j(0)$ selected from  Eq.~\eqref{eq:fullinitcond}. Obtaining the coefficient vectors of $\mathcal I_j(0)$ is straightforward as they are already expressed in terms of the basis elements of the Lie algebra in Eq.~\eqref{eq:transLie}. 

The propagators $U_G$, $U_C$ and $U_D$ considered in our work are of the common form
\be
U_T=e^{-i\theta \mathcal{H}} \ ,
\ee
where $\theta$ is a phase, and $\mathcal{H}$ a linear combination of the basis elements of the Lie algebra. The final condition $\mathcal I_{T j}\equiv U_T \mathcal I_j(0) U_T^{\dagger}$ corresponds to propagating $\mathcal I_j(0)$ with the Hamiltonian $H(t)$ (Eq.~\eqref{eq:ham}) for a duration $\theta$. Starting with the coefficient vector $\bm{a}(0)$ of $\mathcal{I}_j(0)$, one propagates the equation
\be
\dot{\bm{a}}(t)=K \bm{a}(t) \ , 
\ee
where $K$ is the adjoint representation matrix of $H(t)$, for the duration $T_P=\theta$. Then, $\bm{a}(T_P)$ is the coefficient vector of the target 
$\mathcal I_{T j}$.

\subsection{Computing infidelity and gradient}

Piece-wise control is used for the individual $X_j$ and $Z_j$ drivings of the system Hamiltonian (Eq.~\eqref{eq:ham}), while the coupling strength $g$ in the interaction term $X_jX_{j+1}$ is assumed to be constant. The pulse duration $T$ is divided into $M$ intervals, each with a fixed driving amplitude. The optimization process aims to minimize the infidelity (Eq.~\eqref{eq:J} or Eq.~\eqref{eq:fidgate}) by optimizing the set of control amplitudes $\{\{f_j^{l},w_1^l,w_n^l\} : 1\leq l\leq M, 1\leq j\leq n\}$ and the total duration $T$. 
To efficiently evaluate the infidelity during optimization, the operator basis elements $\{ \mathfrak a_j\}$ are normalized, \textit{i.e.} satisfy $\mbox{tr}(\mathfrak{a}_j\mathfrak{a}_k)=\delta_{jk}$. This normalization simplifies the propagation of operators according to Eq.~\eqref{dynamics_aj} and the subsequent calculation of the overlap between the target coefficient vector $\bm{a}_T$ and the final coefficient vector $\bm{a}(T)$ obtained from an initial condition $\bm{a}(0)$.

The control problem is formulated within the Lie algebra as finding the final coefficient vector $\bm{a}(T)$ to minimize the invariant infidelity $\mathcal{J}=1- \bm{a}_T \cdot \bm{a}(T)$, where $\bm{a}_T$ is the target coefficient vector, starting from an initial condition $\bm{a}(0)$. The evolution of $\bm{a}(t)$ is governed by the adjoint representation matrix $K(t)$ of the system Hamiltonian $H(t)$. If one expresses $H(t)=H_0+\sum_k c_k(t) H^c_k$, where $c_k(t)$ is the amplitude of the control operator $H^c_k$, then $K(t)=K_0 + \sum_k c_k(t) K^c_k$ where $K_0$ and $K^c_k$ are the adjoint representations of $H_0$ and $H^c_k$, respectively (see Eq.~\eqref{eq:Kmatrix2}). For the  piece-wise control defined above, the solution of $\dot{\bm{a}}(t)= K(t) \bm{a}(t)$ is 
\be
\bm{a}(T)=\prod_{l=M}^1\mathcal{U}_l\bm{a}(0) \ ,
\ee
where
\be
\mathcal{U}_l=e^{\Delta t\left( K_0+\sum_k c_k^l K_k^c \right)} \ .
\ee
Note that $\mathcal{U}_l$ is unitary as $K(t)$ is anti-hermitian.  One first computes and stores the set of forward and backward propagated vectors, defined by
\begin{align}
\bm{a}_m&=\mathcal{U}_m\mathcal{U}_{m-1}\dots \mathcal{U}_1\bm{a}(0) \ , \nonumber \\
\bm{b}^{\dagger}_{m}&=\bm{a}_T^{\dagger}\mathcal{U}_M\mathcal{U}_{M-1}\dots \mathcal{U}_{m} \ , \nonumber
\end{align}
using the recursive relations $\bm{a}_{m}=\mathcal{U}_{m}\bm{a}_{m-1}$ and $\bm{b}^{\dagger}_m=\bm{b}^{\dagger}_{m+1}\mathcal{U}_m$, then the gradient of $\mathcal{J}$ with respect to $c^l_k$ is
\be
\frac{\partial \mathcal{J}}{\partial c_k^l}=\frac{\partial}{\partial c_k^l}\left(1-\bm{b}^{\dagger}_{l+1}U_l\bm{a}_{l-1}\right)=-\bm{b}^{\dagger}_{l+1}\frac{\partial U_l}{\partial c_k^l}\bm{a}_{l-1} \ .
\ee
The derivative of $\mathcal{U}_l$ is approximated by a 2nd-order method \cite{de2011second}
\be
\frac{\partial U_l}{\partial c_k^l}=\left\{\Delta t K^c_k -\frac{\Delta t^2}{2}\left[K_l,K^c_k \right]\right\}U_l+O(\Delta t^3) \ ,
\ee
where $K_l=K_0+\sum_k c_k^l K^c_k$, thus
\be
\frac{\partial \mathcal{J}}{\partial c_k^l}=-\bm{b}^{\dagger}_{l+1}\left\{ \Delta t K^c_k -\frac{\Delta t^2}{2}\left[K_l,K^c_k \right]\right\}\bm{a}_{l} \ ,
\ee
which is computed by matrix-vector multiplication. The most computationally expensive part is the multiplication of a matrix exponential with a vector for obtaining $\bm{a}_m$ and $\bm{b}^{\dagger}_m$, but it can be done efficiently with the Krylov subspace method owing to the sparsity of $K(t)$. 

For calculating the derivative of $\mathcal{J}$ with respect to the duration $T$, one simply calculates $\mathcal{J}$ again for a perturbed duration $T+dT$ while keeping the amplitudes $c_k^l$ the same, so that
\be
\frac{\partial \mathcal{J}}{\partial T}\approx \frac{\mathcal{J}(T+dT)-\mathcal{J}(T)}{dT} \ .
\ee
In our calculation, we choose $g\, dT=10^{-10}$.

\subsection{Optimisation}\label{sec:optm}
 
We feed the infidelity and its gradient with respect to the control variables to a gradient-based optimisation algorithm.  The interior point method is employed, and the Hessian is approximated with the L-BFGS method. This algorithm is implemented in the fmincon function in Matlab.

For large $n$, there are many shallow local minima and  plateaus in the control landscape for $U_G$ due to the difficulty of reaching the $n$-body interaction term. To systematically address these challenges, a three-step procedure is employed. Initially, the symmetry of the control problem is imposed on the control variables, followed by the relaxation of this symmetry. Specifically, we begin with 
\begin{align}
\nonumber H^c_1(t)&=g\sum_{j=1}^{n-1}X_{j}X_{j+1} +f_1(t)\sum_{j=1}^n Z_j +f_2(t)(Z_1+Z_n)\\
&+w(t)(X_1+X_n) \ ,
\end{align}
which is a translationally invariant control, except at the two ends, \textit{i.e.}, it has the same symmetry as the control targets. The optimisation is done sequentially from $n=5$ to $n=50$ in step of 5 where the optimal control for $n-5$ is used as the initial guess for $n$.

We then lower the infidelity for each $n$ by optimising with a system Hamiltonian with inversion symmetry
\begin{align}
\nonumber H^c_2(t) &= g\sum_{j=1}^{n-1}X_{j}X_{j+1} + \sum_{j=1}^{\lfloor(n+1)/2 \rfloor} f_j(t) (Z_j+Z_{n+1-j}) \\
&+w(t)(X_1+X_n) \ ,
\end{align}
using the final control in the previous step as the initial guess. Finally, we refine the solution further by optimizing with a system Hamiltonian that lacks any symmetry
\begin{align}
H^c_3(t)=g\!\sum_{j=1}^{n-1}X_{j}X_{j+1}\!+\!\sum_{j=1}^n f_j(t) Z_j\!+\!\!\!\!\sum_{j\in\{1,n\}}\!\!\!\!w_j(t)X_j \ .
\end{align}
Each optimisation step is done with a maximum of 1000 function-and-gradient evaluations. At every step, small random perturbations are applied to the initial guess to explore $12$ nearby solutions in parallel, facilitating escape from traps. In many cases the threshold infidelity is reached already with $H^c_1(t)$ or $H^c_2(t)$. 

$T \approx n\pi/2 g$ for the optimisations of $H_G, H_C$, $T\approx n\pi/ g$ for $H_D$,  and the number of time bins in the piece-wise control pulse is $M=10n$ for all three cases. For $U_C$ and $U_D$, $T\approx 2\pi g$ and  $M=10n$. For $U_G$, $T\approx 2n\pi/g$ and $M=20n$.

\subsection{Adiabatic error estimate}

The control target $\mathcal I_D$ for $H_D$ is defined as the operator $\mathcal I(T_a)$ resulting from simulated adiabatic evolution with a very long duration ($T_a=2500 \tau_g$) from the initial operator $\mathcal I(0)$, as described in Sec.~\ref{sec:adb}. In a perfect adiabatic evolution, each eigenstate of ${\cal I}(0)$ would evolve to a corresponding eigenstate of $H_T$ and, thus, ${\cal I}(T_a)$ would commute with $H_T$. To verify this commutativity, one can propagate ${\cal I}(t)$ with $H_T$ for an additional time $T_b$ randomly chosen from the interval $[50 \tau_g, 100 \tau_g]$. In the ideal adiabatic limit, since ${\cal I}(T_a)$ commutes with $H_T$, further evolution would not change the system, resulting in ${\cal I}(T_a+T_b)={\cal I}(T_a)$. Any deviation from this equality is a measure of adiabatic error, and the infidelity $1-\mbox{tr}({\cal I}(T_a){\cal I}(T_a+T_b))/\mbox{tr}({\cal I}(T_a)^2)$ can be used as an estimate of this error.

\subsection{MPS verification}
The state infidelity for $H_D$ in Fig.~\ref{fig:infid} was obtained with MPS. The ground state of $H_D$ is computed with DMRG, and the final state is obtained by evolving the initial state $\ket{\Psi_{0}}$ with the MPO (matrix product operator) $\text{W}^{II}$ method \cite{Zaletel2015}, where each time bin is further divided into $20$ smaller steps to reduce error. All MPS calculations are done using the Tenpy package \cite{tenpy}.

\section{Relation between the state infidelity bound and the operator infidelity}\label{sec:bound}

Given the system Hamiltonian $H(t)$ (Eq.~\eqref{eq:ham}), the initial condition ${\cal I}(0)=\sum_j Z_j$ and a control target ${\cal I}_T$ that coincides with the target Hamiltonian $H_T$ (which includes the cases of $H_G$ and $H_C$ defined in Sec.~\ref{sec:targethamiltonians}),
the bound
\be
\mathcal{B}=\frac{\bra{\Psi(T)}H_T\ket{\Psi(T)}-E_0}{E_1-E_0}
\label{eq:B}
\ee
in the state infidelity defined in Eq.~
\eqref{bound_infidelity}
depends on the operator infidelity $\mathcal{J}$ via the linear relation 
\be
\mathcal B=\frac{n\mathcal{J}-s}{E_1-E_0} \ ,
\label{eq:Jtoprove}
\ee
with a term $s$ that, under assumptions on typicality, becomes vanishingly small in the limit $n\to\infty$.

This is proven in the following.

Given the requirement that ${\cal I}_T=H_T$, the expectation value in Eq.~\eqref{eq:B} satisfies the equality \be
\bra{\Psi(T)}H_T\ket{\Psi(T)}
=
\bra{\Psi(0)}\tilde{\mathcal{I}}(T)\ket{\Psi(0)}\ ,
\ee
where
$
\tilde{\mathcal{I}}(T)=U^{\dagger}(T)\mathcal{I}_T U(T)
$,
 $\ket{\Psi(0)}$ is the ground state of $\mathcal{I}(0)$, and $E_0$ the ground energy of both $H_T$ and $\mathcal{I}(0)$.
Using the relations
\be
\mbox{tr}(\mathcal{I}(T)\mathcal{I}_T)=\mbox{tr}(U(T)\mathcal{I}(0)U^{\dagger}(T)\mathcal{I}_T)=\mbox{tr}(\mathcal{I}(0)\tilde{\mathcal{I}}(T))
\ee
and $\mbox{tr}(\mathcal{I}_T^2)=\mbox{tr}(\mathcal{I}^2(0))$, the operator infidelity in Eq.~\eqref{eq:J} can be expressed as
\be
\mathcal{J}=1-\frac{\mbox{tr}(\mathcal{I}(0)\tilde{\mathcal{I}}(T))}{\mbox{tr}(\mathcal{I}^2(0))} \ . 
\ee

If the control target is reached exactly, the identity $\tilde{\mathcal{I}}(T)=\mathcal{I}(0)$ holds,
and any imperfection of the optimal control solution can be captured by the deviation 
\be
\Delta \mathcal{I}=\mathcal{I}(0)-\tilde{\mathcal{I}}(T)\ .
\label{eq:tildeI}
\ee

The deviation $\Delta \mathcal{I}$ can be expanded in terms of the operators $\mathfrak{a}_j$ (Eqs.~\eqref{eq:transLie}) as $\Delta \mathcal{I}=\sum_j c(\mathfrak{a}_j)\mathfrak{a}_j$,
with expansion coefficients
\be
c(\mathfrak{a}_j)=\frac{\tr(\mathfrak{a}_j\ \Delta\mathcal{I})}{\tr(\mathfrak{a}_j^2)}\ .
\label{eq:expcoeff}
\ee

The operator infidelity can be expressed as
\be
\mathcal{J}=\frac{\mbox{tr}(\mathcal{I}(0)\Delta \mathcal{I})}{\mbox{tr}(\mathcal{I}^2(0))} \ .
\ee

With $\mathcal{I}(0)=\sum_j Z_j$, $\mbox{tr}(\mathcal{I}^2(0))=n2^n$, and the overlaps
$\mbox{tr}(\mathcal{I}(0)Z_j)=2^n$,
and $\mbox{tr}(\mathcal{I}(0)\mathfrak{a}_j)=0$ for all operators $\mathfrak{a}_j$
that are not single-spin $Z$ operators ({\it i.e.} Eqs.~\eqref{eq:transLie2nd} to \eqref{eq:transLieZn}),
the operator infidelity reduces to
\be
\mathcal{J}=\frac{\sum_j c(Z_j)}{n}\ ,
\label{eq:J_zj}
\ee
in terms of the expansion coefficients $c(Z_j)$ defined in Eq.~\eqref{eq:expcoeff}.

Given the ground state energy $\bra{\Psi(0)}\mathcal{I}(0)\ket{\Psi(0)}=E_0$, the bound ${\cal B}$ can be expressed as
\be
\mathcal B =-\frac{\bra{\Psi(0)}\Delta \mathcal{I}\ket{\Psi(0)}}{E_1-E_0}\ .
\ee
With $\Delta \mathcal{I}$ expanded in terms of the operators $\mathfrak{a}_j$ (Eqs.~(\eqref{eq:transLie}) and the expectation values
$\bra{\Psi(0)}Z_j\ket{\Psi(0)}=-1$,
$\bra{\Psi(0)}\prod_j Z_j\ket{\Psi(0)}=(-1)^n$,
and
$\bra{\Psi(0)}\mathfrak{a}_j\ket{\Psi(0)}=0$ for all other cases ({\it i.e.} Eqs.~\eqref{eq:transLie2nd} to \eqref{eq:transLie2ndlast}), this yields
\be
\mathcal B=\frac{\sum_j c(Z_j) - s}{E_1-E_0} \ ,
\label{eq:B_with_s}
\ee
where $s=(-1)^n c(\prod_j Z_j)$, and the expansion coefficient $c(\prod_j Z_j)$ is defined in Eq.~\eqref{eq:expcoeff}.
Together with Eq.~\eqref{eq:J_zj}, this results in Eq.~\eqref{eq:Jtoprove}. \\

While it is conceivable that $\Delta {\cal I}$ is proportional to the operator $\prod_j Z_j$, one would expect that, typically, $\Delta {\cal I}$ would have contributions from all operators $\mathfrak{a}_j$ in Eq.~\eqref{eq:transLie}.
In estimating the magnitude of $s$, we will thus assume the expansion coefficients $c(\mathfrak{a}_j)$ to be random with each coefficient resulting from a distribution with zero mean and standard deviation $\delta$.
The typical magnitude of the term \be
n\mathcal{J}=\sum_{j=1}^n c(Z_j)
\ee
is then given by its standard deviation, {\it i.e.} by $\sqrt{n}\delta$.
The typical magnitude of $c\left(
\prod_j Z_j\right)$, on the other hand, is given by $\delta$.
The term $s$ thus becomes negligible as compared to the term $n\mathcal{J}$ for large systems.

\begin{figure}[h!]
\centering
\includegraphics[width=0.9\columnwidth]{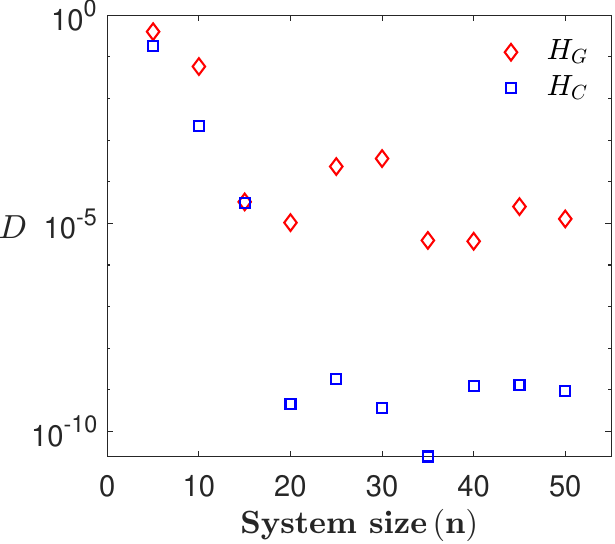}
\caption{
Deviation $D=\left\vert 1-(E_1-E_0)\mathcal B/n\mathcal J\right \vert=\vert s \vert/n\mathcal J$, in log-scale, from the linear relation between the bound $\mathcal{B}$ on state infidelity and operator infidelity $\mathcal J$ that is expected to hold in the limit $n\to\infty$ following Eq.~\eqref{eq:Jtoprove}.
The deviation obtained from the data shown in Fig.~\ref{fig:infid}a decays rapidly with $n$, consistently with Eq.~\eqref{eq:Jtoprove}.
}\label{fig:bound}
\end{figure}

Fig.~\ref{fig:bound} depicts the deviation of the actual results of Fig.~\ref{fig:infid}a from the linear dependence $\mathcal{B} = \frac{n \mathcal{J}}{E_1-E_0}$ as a function of the system size $n$.
The figure shows that the deviation decays rapidly with $n$. The actual results thus follow the linear dependence in good approximation.

\section{Lie algebra of the hexagonal spin ladder}\label{lie_hexagonal}

As shown in the following,
the Lie algebra generated by the operators in Eq.~\eqref{eq:Hhoney_mod} for the spin ladder depicted in Fig.~\ref{fig:2D_4} scales as $(25n^2-40n+12)/32$ in the number of spins $n=4n_h+2$ with $n_h$ hexagons.

In order to arrive at this scaling, it is convenient to start with two non-interacting chains.

The Lie algebra for the top chain is generated by Eqs.~\eqref{eq:Hhoney_topYY} and \eqref{eq:Hhoney_topZZ}. With the short-hand notation $X_{jk}=\prod_{i=1}^{k}X_{i+j}^1$, and
\begin{subequations}
\bqa
    \overline{YZ}_{jk}  &=& Y_{j}^1X_{jk}Z_{j+k+1}^1 \ , \\
    \overline{ZY}_{jk}  &=& Z_{j}^1X_{jk}Y_{j+k+1}^1 \ , \\
    \overline{YY}_{jk}  &=& Y_{j}^1X_{jk}Y_{j+k+1}^1 \ , \\
    \overline{ZZ}_{jk}  &=& Z_{j}^1X_{jk}Z_{j+k+1}^1 \ ,
\eqa
\end{subequations}
it is given by the operators $\gamma_{jk}$ with $j\in[1,m-1]$ and $k\in[0,m-j-1]$,
where
\begin{align}
\gamma_{jk}=
    \begin{cases}
    \overline{YZ}_{jk} \ , \quad \text{for odd } j \text{ and odd } k \ , \\
    \overline{YY}_{jk}\ , \quad \text{for odd } j \text{ and even } k \ , \\
    \overline{ZY}_{jk}\ , \quad \text{for even } j \text{ and odd } k \ , \\
    \overline{ZZ}_{jk}\ , \quad \text{for even } j \text{ and even } k \ .
    \end{cases}
\label{eq:Hhoney_mod_ope}
\end{align}

The Lie algebra for the bottom chain is generated by Eqs.~\eqref{eq:Hhoney_bottomXX} and \eqref{eq:Hhoney_bottomZZ}, and it includes (besides Eqs.~\eqref{eq:Hhoney_bottomXX} and \eqref{eq:Hhoney_bottomZZ})
\begin{align}
X_{2j}^2Y_{2j+1}^2X_{2j+2}^2 \ , 
\label{eq:lower_chain}
\end{align}
with $j\in[1,(m-3)/2]$. 

With  $s\in \{-1,1\}$, $j\in[1,(m-1)/2]$ , $k\in [0, m-j-1]$, and with $\Gamma_{2j}$ adopting the four different values
\bqa
\Gamma_{2j}&=&
\begin{cases}
    X_{2j-1}^2X_{2j}^2 \ , \\
    Y_{2j-1}^2\mathbbm{1}_{2j}^2 \ ,
\end{cases}
\eqa
 for $s=-1$, 
\bqa
\Gamma_{2j}&=&
\begin{cases}
    X_{2j}^2Z_{2j+1}^2  \ , \\
    \mathbbm{1}_{2j}^2Y_{2j+1}^2 \ ,
\end{cases}
\eqa
for $s=1$, the Lie algebra generated by the full set of operators in Eq.~\eqref{eq:Hhoney_mod} contains the additional terms

\begin{subequations} 
\begin{flalign}
    &X_{2j+s}^1 \Gamma_{2j} \ , \\
    &\overline{ZY}_{2j+s,k} \Gamma_{2j} \ , \\
    &\overline{ZY}_{2j-1-k+s,k} \Gamma_{2j} \ , 
\end{flalign}
 for even  $k$, and
 \begin{flalign}
    &\overline{ZZ}_{2j+s,k} \Gamma_{2j} \ , \\
    &\overline{ZY}_{2j+s,k} \Gamma_{2j} Y_{2j+k+s+1}^2 \ , \\
    &\overline{ZY}_{2j+s,k} \Gamma_{2j} X_{2j+k+s}^2Z_{2j+k+s+1}^2 \ ,  \\ &\overline{ZY}_{2j+s,k} \Gamma_{2j} X_{2j+k+s+1}^2Z_{2j+k+s+2}^2 \ , \\
    &\overline{YY}_{2j-1-k+s,k} \Gamma_{2j} \ ,  \\
    &\overline{ZY}_{2j-1-k+s,k} Y_{2j-1-k+s}^2\Gamma_{2j} \ , 
    \label{eq:origin_plaq}\\
    &\overline{ZY}_{2j-1-k+s,k} X_{2j-2-k+s}^2Z_{2j-1-k+s}^2\Gamma_{2j} \ ,  \\
    &\overline{ZY}_{2j-1-k+s,k} X_{2j-1-k+s}^2X_{2j-k+s}^2\Gamma_{2j} \ , 
\end{flalign}
    for odd $k$.
\label{eq:Hhoney_mod_ope2}
\end{subequations} 

Given that Eqs.~\eqref{eq:Hhoney_bottomXX}-\eqref{eq:Hhoney_bottomZZ} contain $(n-2)/2$ elements, Eq.~\eqref{eq:Hhoney_mod_ope} includes $(n^2/8 - n/4)$ elements (encompassing Eqs.~\eqref{eq:Hhoney_topYY} and \eqref{eq:Hhoney_topZZ}),  Eq.~\eqref{eq:lower_chain} contributes with $(n-6)/4$ elements, and Eq.~\eqref{eq:Hhoney_mod_ope2} adds $(21n^2-56n+92)/32$ operators, the size of the Lie algebra grows as $(25n^2-40n+12)/32$ with the number of qubits $n=2 + 4n_h$ defined in terms of the number of hexagons $n_h$. 

For $k=1$, $s=1$ and $\Gamma_{2j}=X_{2j}^2Z_{2j+1}^2$, the operator in Eq.~\eqref{eq:origin_plaq} reduces to the $6$-body plaquette operators $P_j= Z_{2j-1}^1X_{2j}^1Y_{2j+1}^1Y_{2j-1}^2X_{2j}^2Z_{2j+1}^2$ in Eq.~\eqref{eq:plaquette_op}.

\bibliography{Lie.bib}

\end{document}